\documentclass[useAMS,usenatbib]{mn2e}
\usepackage{graphicx}

\def \aj {AJ}
\def \mnras {MNRAS}
\def \pasp {PASP}
\def \apj {ApJ}
\def \apjs {ApJS}

\def \aap {A\&A}
\def \nat {Nature}
\def \araa {ARAA}

\def\lesssim{\mathrel{\hbox{\rlap{\hbox{\lower4pt\hbox{$\sim$}}}\hbox{$<$}}}}
\def\gtrsim{\mathrel{\hbox{\rlap{\hbox{\lower4pt\hbox{$\sim$}}}\hbox{$>$}}}}
\title[The Type Ibn SNe 2010al and 2011hw]{Massive stars exploding in a He-rich circumstellar medium. IV. Transitional Type Ibn Supernovae}
\author[Pastorello et al.]{A. Pastorello,$^1$\thanks{andrea.pastorello@oapd.inaf.it}
S. Benetti,$^1$ P. J. Brown,$^2$  D. Y. Tsvetkov,$^3$   C. Inserra,$^4$ 
\newauthor S. Taubenberger,$^6$ L. Tomasella,$^1$ M. Fraser,$^5$  D. J. Rich,$^7$ M. T. Botticella,$^8$
\newauthor  F. Bufano,$^9$ E. Cappellaro,$^1$  M. Ergon,$^{10}$ E. S. Gorbovskoy,$^{3,11}$ A. Harutyunyan,$^{12}$ 
\newauthor F. Huang,$^{13,14}$   R. Kotak,$^4$  V. M. Lipunov,$^{3,11}$  L. Magill,$^4$ M. Miluzio,$^{1,15}$ N. Morrell,$^{16}$ 
\newauthor   P. Ochner,$^1$ S. J. Smartt,$^4$  J. Sollerman,$^{10}$ S. Spiro,$^{1,17}$ M. D. Stritzinger,$^{18}$
\newauthor    M. Turatto,$^1$ S. Valenti,$^{19,20}$ X. Wang,$^{14}$  D. E. Wright,$^4$ V. V. Yurkov,$^{21}$
\newauthor   L. Zampieri,$^{1}$ and T. Zhang.$^{22}$ \\
\\
$^{1}$INAF-Osservatorio Astronomico di Padova, Vicolo dell'Osservatorio 5,  35122 Padova, Italy\\
$^{2}$George P. and Cynthia Woods Mitchell Institute for Fundamental Physics \& Astronomy, Texas A. \& M. University, \\Department of Physics and Astronomy, 4242 TAMU, College Station, TX 77843, USA\\
$^{3}$Sternberg Astronomical Institute of Lomonosov Moscow State University, University Avenue 13, 119992 Moscow, Russia\\
$^{4}$Astrophysics Research Centre, School of Mathematics and Physics, Queen's University Belfast, Belfast BT7 1NN, United Kingdom\\
$^{5}$Institute of Astronomy, University of Cambridge, Madingley Road, Cambridge, CB3 0HA, United Kingdom\\
$^{6}$Max-Planck-Institut f\"ur Astrophysik, Karl-Schwarzschild-Str. 1, 85741 Garching, Germany\\
$^{7}$Rich Observatory, 62 Wessnette Dr., Hampden, ME, USA\\
$^{8}$INAF - Osservatorio Astronomico d Capodimonte, Salita Moiariello 16, 80131 Napoli, Italy\\
$^{9}$Departamento de Ciencias Fisicas, Universidad Andres Bello, Avda. Republica 252, Santiago, Santiago RM, Chile\\
$^{10}$Department of Astronomy, The Oskar Klein Centre, Stockholm University, 106 91 Stockholm, Sweden\\
$^{11}$Lomonosov Moscow State University, GSP-1, Leninskie Gory, Moscow, 119991, Russia\\ 
$^{12}$Fundaci\'on Galileo Galilei-INAF, Telescopio Nazionale Galileo, Rambla JosŽ Ana Fern\'andez P\'erez 7, 38712 Bre\~na Baja, TF, Spain\\
$^{13}$Department of Astronomy, Beijing Normal University, Beijing, 100875, China\\
$^{14}$Physics Department and Tsinghua Center for Astrophysics, Tsinghua University, Beijing, 100084, China\\
$^{15}$Dipartimento di Astronomia, Universit\'a di Padova, Vicolo dellÕOsservatorio 3, 35122 Padova, Italy\\
$^{16}$Carnegie Observatories, Las Campanas Observatory, Colina El Pino, Casilla 601, Chile \\
$^{17}$Department of Physics (Astrophysics), University of Oxford, DWB, Keble Road, Oxford OX1 3RH, United Kingdom\\
$^{18}$Department of Physics and Astronomy, Aarhus University, Ny Munkegade, DK-8000 Aarhus C, Denmark \\
$^{19}$Las Cumbres Observatory Global Telescope Network, Inc. Santa Barbara, CA 93117, USA\\
$^{20}$Department of Physics, University of California Santa Barbara, Santa Barbara, CA 93106-9530, USA\\
$^{21}$Blagoveshchensk State Pedagogical University, ul. Lenina 104, 675000 Blagoveshchensk, Russia\\
$^{22}$National Astronomical Observatory of China, Chinese Academy of Sciences, Beijing, 100012, China} 

\begin{document}

\date{Accepted 20XX Month XX. Received 20XX Month XX; in original form 20XX Month XX}

\pagerange{\pageref{firstpage}--\pageref{lastpage}} \pubyear{2014}

\label{firstpage}

\maketitle

\clearpage
\begin{abstract}
We present ultraviolet, optical and near-infrared data of the Type Ibn supernovae (SNe) 2010al and 2011hw.
SN 2010al reaches an absolute magnitude at peak of $M_R$ = $-$18.86 $\pm$ 0.21.
Its early light curve shows similarities with normal SNe Ib, with a rise to maximum slower than  most
SNe Ibn. The spectra are dominated by a blue continuum at early stages, with narrow 
P-Cygni He I lines indicating the presence of a slow-moving, He-rich circumstellar medium.
At later epochs the spectra well match those of the prototypical SN Ibn 2006jc, although the broader lines
suggest that a significant amount of He was still present in the stellar envelope at the time
of the explosion. SN 2011hw is somewhat different. It was discovered after the first maximum, but the light curve
shows a double-peak. The absolute magnitude at discovery is similar to that of the 
second peak ($M_R$ = $-$18.59 $\pm$ 0.25), and slightly fainter than the average of SNe Ibn.
Though the spectra of SN 2011hw are similar to those of SN 2006jc, coronal lines and narrow Balmer lines
are cleary detected. This indicates substantial interaction of the SN ejecta with He-rich, but not H-free, circumstellar material.
The spectra of SN 2011hw suggest that it is a transitional SN Ibn/IIn event similar to SN 2005la. 
While for SN 2010al the spectro-photometric evolution favours a H-deprived Wolf-Rayet progenitor (of WN-type), we agree with the   
conclusion of \citet{smi12} that the precursor of SN 2011hw was likely in transition from
a luminous blue variable to an early Wolf-Rayet (Ofpe/WN9) stage.
\end{abstract}

\section{Introduction} \label{intro}

Type Ibn supernovae (SNe) are a poorly understood family of core-collapse SNe (CC SNe). The label
``SNe Ibn'' was introduced a few years ago \citep{pasto08a}, although the first object of this class was discovered
almost a decade before \citep[SN 1999cq, ][]{mat00}.
Early spectra of SNe Ibn are blue, and show simultaneously broad lines 
of intermediate mass elements (IME) and relatively narrow lines of He, while lines of H are weak
or absent. The lack of narrow Balmer lines is an observational  property distinguishing this
sub-group from classical Type IIn SNe.
Although SNe Ibn are  quite luminous, their photometric evolution is usually
very rapid, with a very fast rise to maximum and a fast post-peak decline, at least in the optical 
 bands. The latter has been proposed to be a signature of relatively early dust formation in a cool dense shell 
formed in the post-shock circumstellar medium \citep[CSM; see e.g.][]{smi08,seppo08}. SNe Ibn are generally interpreted as Type Ib/c SN 
explosions occurring within a He-rich circumstellar environment. 

The prototype of Type Ibn events is SN 2006jc, the first CC SN observed to explode a short time
(2 yrs) after a major eruptive episode of the progenitor registered by the amateur astronomer K. Itagaki \citep{pasto07}. 
Unfortunately SN 2006jc was discovered
a few weeks after explosion, and we could not follow the early-time evolution.
Nevertheless, whilst high-quality data are available in the literature for 
SN 2006jc\footnote{Multi-wavelength observations of SN 2006jc have been presented by  
\citet{pasto07,fol07,imm08,smi08,pasto08a,seppo08,tom08,elisa08,noz08,sak09,anu09,buf09,mod14,bia14}.}, 
for other Type Ibn SNe only sparse optical data have been published
\citep[e.g.][]{mat00,pasto08a,pasto08b}. 

Among recent additions to this family, remarkable objects are: 
 PS1-12sk, iPFT13beo, LSQ12btw, LSQ13ccw and OGLE-2012-SN-006. 
PS1-12sk was discovered by the Pan-STARRS1 survey \citep{kai02} and has been proposed to be the first Type Ibn SN hosted in an
elliptical galaxy \citep{san13}. iPTF13beo is an intermediate Palomar Transient Factory \citep[iPTF,][]{kul13} discovery. It was
detected soon after the explosion, and it showed a sort of double-peaked
light curve \citep{gor14}.
LSQ12btw and LSQ13ccw are two objects discovered by the la Silla-Quest survey \citep{rab11} and classified by the
``Public ESO Spectroscopic Survey of Transient Objects'' \citep[PESSTO,][]{val12,kar13}.
The data of these two transients are presented in \citet{pasto13}. 
OGLE-2012-SN-006 was discovered by the OGLE IV survey \citep{wyr12} and classified as a Type Ibn SN a few months later \citep{pri13}. 
This is the first SN Ibn with a slowly evolving late-time optical
light curve \citep{pasto13b}. 

The first opportunity of a complete monitoring of a Type Ibn SN starting soon after explosion was provided
with the  discovery of SN 2010al in the spiral galaxy UGC 4286 on March 13, 2010 
\citep{rich10}. Early-time spectra showed some resemblance with those of the
 Type IIn SNe 1998S and 2001fa, and with early spectra of the Type II SN 1983K, for the detection of narrow H Balmer lines in emission
and the  presence of narrow features usually found in Wolf-Rayet winds, such as the N III / C III blend at
4640 \AA~and the He II $\lambda$4686 and $\lambda$5412 lines \citep{coo10,max10,sil10}. However, the disappearance of H lines
and the strengthening of He I features at later phases (see Section \ref{spec}) suggest us to revise the classification of SN 2010al
as a Type Ibn event. Two XShooter spectra of SN 2010al have been
presented in \citet{pasto12}. 

More recently, another interesting transient
named SN 2011hw was discovered by B. Dintinjana and H. Mikuz (Crni Vrh Observatory)
and classified by our team \citep{val11}. The object appeared to be similar to the transitional Type-IIn/Ibn SN
2005la \citep{pasto08b} because of the
presence of H lines in emission, though weaker than the most prominent He I emission features
\citep{din11,val11}. Sparse photometry and a nice spectral sequence of SN 2011hw have been presented 
by \citet{smi12}, together with a comprehensive discussion on the nature of this transitional event.

In this paper we will present and analyse new data of the Type Ibn SNe 2010al and 2011hw collected in
the framework of an extensive international collaboration assembled on the
ESO-NTT and TNG {\sl Supernova Variety and Nucleosynthesis Yields} Large 
programs.\footnote{\it  http://sngroup.oapd.inaf.it/esolarge.html}
In Section \ref{obs} we will present the observations of the two SNe. In Section \ref{photo} we will describe the photometric data reduction
techniques and discuss the light curves of SNe 2010al and 2011hw. Spectroscopic data will be illustrated and analysed
in Section \ref{spec}. 
Finally, a discussion and a summary will follow in Section \ref{disc}.

\begin{figure}
   \centering
   \includegraphics[width=3.4in]{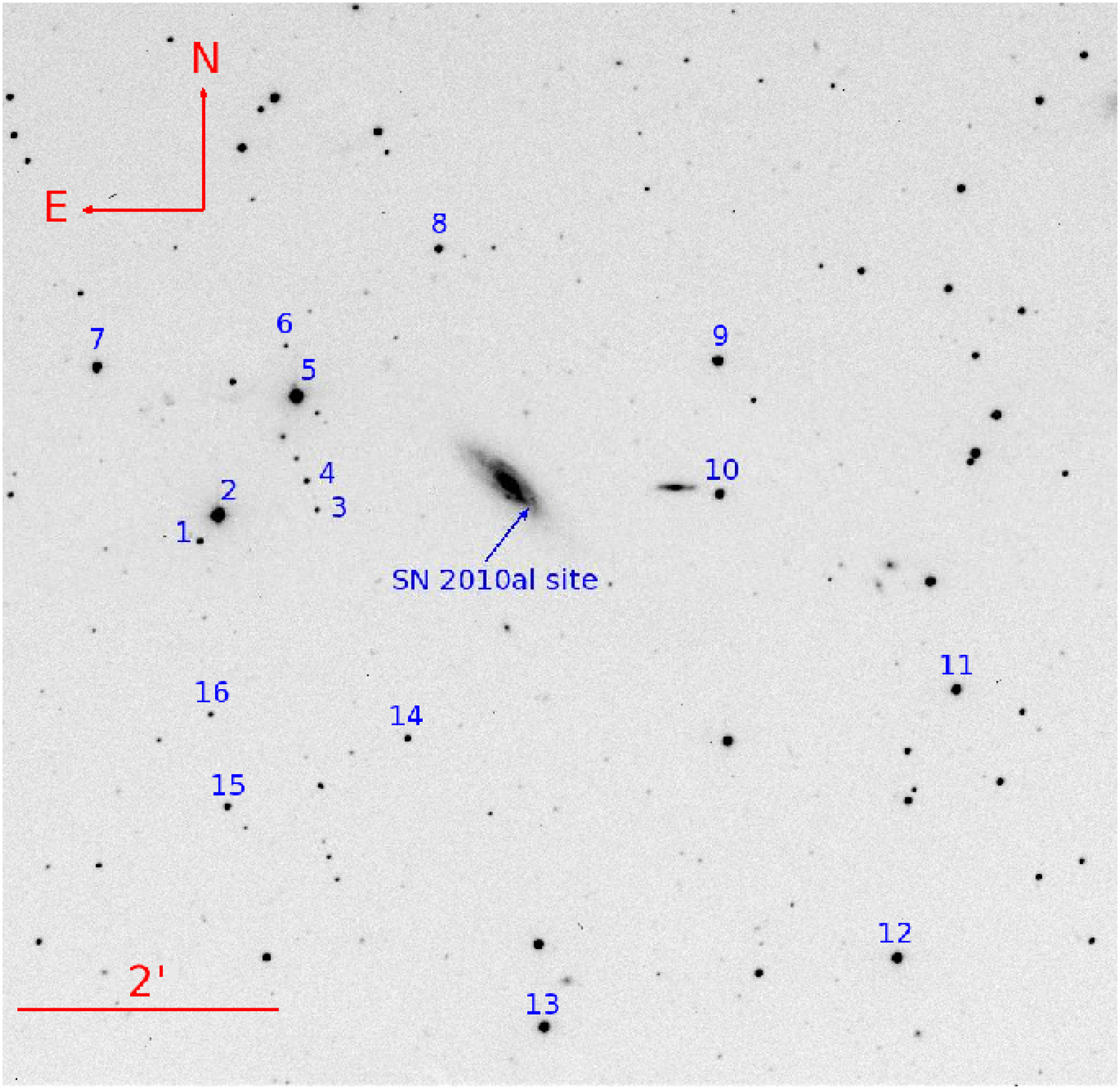} 
 
\medskip
   \includegraphics[width=3.4in]{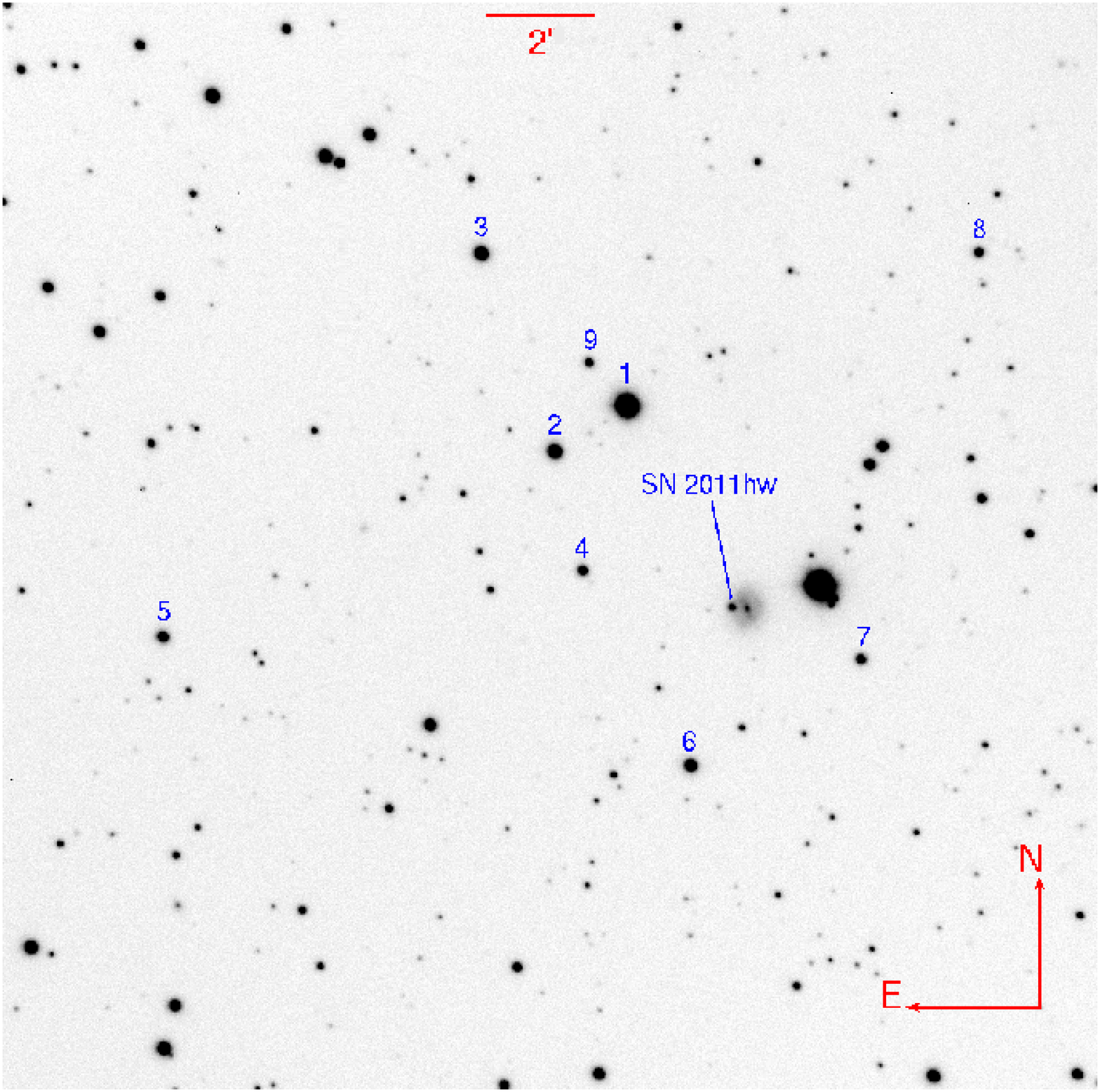} 
   \caption{$R$-band images of the fields of SNe 2010al ({\bf top}) and 2011hw ({\bf bottom)}. The sequence stars used to calibrate the magnitudes
  of the two SNe are marked. 
}
   \label{fig:app_maps}
\end{figure}

\section{Observations}  \label{obs}
We started our optical and near-infrared (NIR) observational campaigns soon after the 
classification announcements of the two SNe, using the instruments available to our collaboration: 
the 8.2-m Very Large Telescope  (VLT - UT2 module) with XShooter (Cerro Paranal, Chile); 
the 3.58-m New Technology Telescope (NTT) equipped with EFOSC2 and SOFI (La Silla, Chile); 
the 1.82-m Copernico Telescope with AFOSC and the 67/92-cm Schmidt Telescope (Mt. Ekar, near Asiago, Italy); 
the 2.2-m telescope in Calar Alto (Almeria, Spain) with CAFOS; 
the two 0.40-m MASTER telescopes\footnote{Information on the genesis and management of the Master Network can be found in \citet{lip10} and \citet{kor12}.}
  in Kislovodsk (Caucasian region, Russia)
and Blagoveschensk (Far East region, Russia) both equipped with Apogee Alta U16M CCDs;
the 2.0-m Faulkes North telescope with the EM01 camera (Haleakala, Hawaii Islands, USA);
the 0.80-m Tsinghua-NAOC (National Astronomical Observatories of China) Telescope (Xinglong Observatory, Yanshan mountains, Hebei, China), equipped with 
a Princeton Instruments VersArray:1300B CCD; 
the 3.58-m Telescopio Nazionale Galileo (TNG)  with Dolores and NICS;
the 4.2-m William Herschel Telescope (WHT) equipped with ACAM and ISIS;
the 2.0-m Liverpool Telescope (LT) with RATCam and SupIRCam; 
the 2.56m Nordic Optical Telescope (NOT) with ALFOSC (La Palma, Canary Islands, Spain).
Additional photometry with small-size telescopes was kindly provided by amateur astronomers.
Both SNe 2010al and 2011hw were visible for only 60-70 days after their discoveries, then they disappeared behind the Sun.
We tried to recover SN 2010al at very late phases, but it was only visible in NIR observations obtained with the 8.4-m Large Binocular 
Telescope (Mt. Graham, Arizona; USA) equipped with Lucifer.
Additional space observations of SNe 2010al and 2011hw in the ultra-violet (UV) and optical bands were
obtained with the SWIFT satellite and its Ultraviolet/Optical Telescope (UVOT). These data were useful to constrain the large energy
contribution of the UV domain in the early phases of the evolution of the two SNe.\footnote
{SN 2010al was also targeted by the Hubble Space Telescope in the UV domain, at almost the same epochs
as our X-Shooter spectra \citep{kir10}.}

\section{Photometry} \label{photo}
\subsection{Data reduction}  \label{data_redu}

Data reduction was performed following standard prescriptions in IRAF.\footnote{IRAF is distributed by the National Optical 
Astronomy Observatory, which is operated by the Association of Universities for Research in Astronomy (AURA) under cooperative agreement with 
the National Science Foundation.} Original images were first overscan-, bias-, flat field- and fringing-corrected, and then 
the unexposed regions of the images were trimmed. 
Occasionally, in order to increase the signal-to-noise ratio, several subsequent exposures were combined.

In order to remove the contribution of the bright background in the NIR images,
we subtracted from individual science frames adjacent sky images, and then we combined the sky-subtracted SN exposures. 
The sky images were obtained by median-combining 
several dithered exposures of stellar regions in the proximity of the SN location.

The SN magnitudes were measured using a point spread function (PSF) fitting technique. 
Photometric zero-points and colour terms were obtained through 
observations of  standard star fields in the same nights as the SN observations. 
The photometric calibrations in the optical domain were based on the \citet{land92} catalogue. The inferred zero-points allowed us 
to calibrate the magnitudes of local stellar sequences in the fields of the two SNe (cfr. Figure \ref{fig:app_maps} and Table \ref{tab_seqstars}). 
For non-photometric nights, we applied zero-point corrections derived by comparing the magnitudes of the 
local sequence stars of those nights with the average magnitudes obtained using a few photometric nights.
With the corrected zero-points we estimated the final SN apparent magnitudes (Tables \ref{tab_ph_10al} and \ref{tab_ph_11hw}) at all epochs.
The calibration of the NIR photometry of SN 2010al was performed with respect to the 2MASS catalogue magnitudes \citep{skru06} of the same local stellar sequence  
used for the optical photometry, and the final NIR SN magnitudes are also reported in Table \ref{tab_ph_10al}.

\begin{figure*}
   \centering
  \includegraphics[angle=270,width=7.15in]{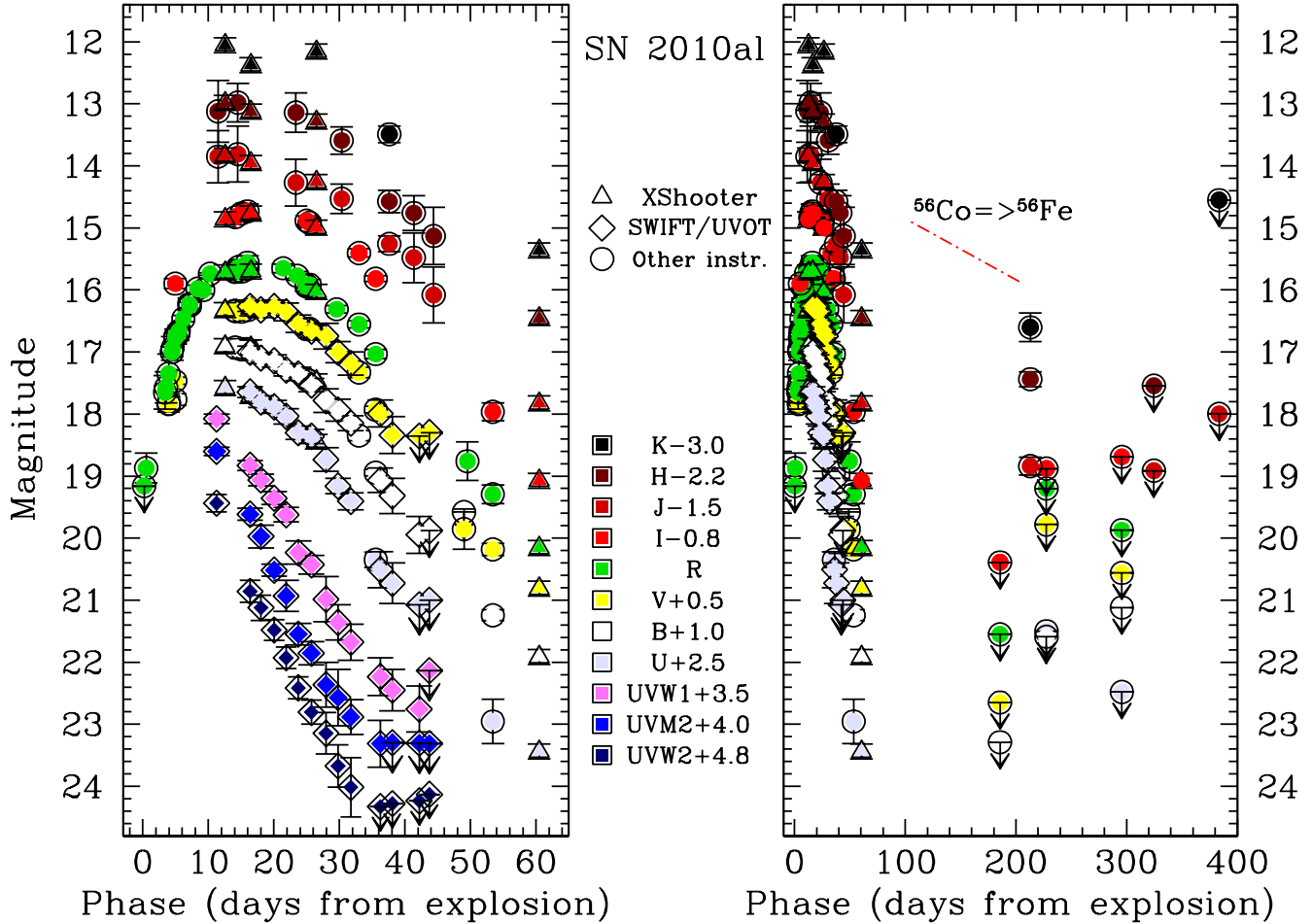}
   \caption{{\bf Left:}  Early-epoch UV (Swift/UVOT) + optical + NIR light curves of SN 2010al. The spectro-photometric points obtained from the
   XShooter spectra have also been included (see Tab. \ref{tab_ph_10al}). Only significant detection limits have been shown in the figure.
   {\bf Right:} optical + NIR light curves of SN 2010al, including  late-time observations. Spectro-photometric observations with XShooter are indicated with triangles, 
SWIFT/UVOT data with diamonds, and data obtained with other instruments with circles. Different colours identify different photometric bands.
Swift/UVOT optical-band magnitudes have been shifted 
by $+$0.082 mag in $U$, $+$0.026 mag in $B$ and $-$0.037 mag in $V$ to match the ground-based  photometry. The light curve decline rate expected
from the $^{56}$Co to $^{56}$Ni decay is also shown to guide the eye.}
   \label{fig:10al_LC}
\end{figure*}

In addition to the ground-based observations,  UV and optical follow-up observations for both SNe were obtained using the Swift
satellite\footnote{Proposal PIs: P. J. Brown; T. Prichard; A. M. Soderberg} equipped with UVOT \citep{rom05,poo08}. 
The data of these two SNe have been presented in \citet{prit13}; here we perform independent measurements
on the same dataset. UVOT photometry was performed following the method detailed in \citet{bro09}.
Since images without the SN were available in the cases of SNe 2010al and 2011hw, the template 
subtraction method was applied to remove the host galaxies and hence improve the photometric measurements (obtained 
using a 3 to 5 arcsec aperture). We note that no template subtraction was applied in the photometry  
presented by \citet{prit13}. Optimised zero-points from \citet{bre11} were then used to convert count rates to the final 
UVOT magnitudes.\footnote{Updated UVOT calibration files (released on January 18, 2013) were
collected from the Swift Calibration Database: {\it http://heasarc.gsfc.nasa.gov/docs/heasarc/caldb/swift/}.} 
The UV magnitudes of a few reference stars in the fields of the two SNe are reported in Table \ref{tab_seqstars_UV}.

\subsection{Distance and reddening estimates}  \label{red_dis}

The location of SN 2010al is $RA = 8^h14^m15^s.91, Dec = +18^o26'18''.2$ (equinox J2000.0), 9.5 arcsec West and 8.1 arcsec South of
the center of UGC 4286, an edge-on Sab-type spiral galaxy \citep[][see also Figure \ref{fig:app_maps}, top]{rich10}. Its recessional 
velocity \footnote{from Hyperleda, {\it http://leda.univ-lyon1.fr/ }} corrected for Virgo infall
is $v_{Vir}$ = 5157 km s$^{-1}$. Adopting as Hubble constant H$_0$ = 73 km s$^{-1}$ Mpc$^{-1}$ ($\Omega_M$ = 0.27, $\Omega_\Lambda$ = 0.73), 
we obtain a luminosity distance of about 71.6 Mpc (corresponding to a distance modulus $\mu$ = 34.27 $\pm$ 0.16 mag). 
The Galactic interstellar absorption is estimated to be $A_B$(MW) = 0.17 mag \citep{sch11}. 
In order to estimate the host galaxy contribution to the total reddening, we inspected our higher resolution XShooter spectra
(see Section \ref{sec:spec10al}). The two lines of the Na I doublet (Na ID) are visible and deblended both at z=0 and at the host galaxy reshift.
We note that the ratio between the equivalent widths (EWs) of the Na I lines of the host galaxy and the Galactic component is 0.43. 
Assuming the same dust to gas ratio and similar dust properties in the Galaxy and in UGC 4286,
we can reasonably estimate  that the host galaxy contribution to the reddening is lower (by a 
factor 0.43) than the Galactic contribution, i.e. $A_B$(host) = 0.07 mag. 
Therefore, hereafter, we will adopt for SN 2010al a total line-of-sight extinction of $A_B$(tot) = 0.24 mag in the $B$ band.

SN 2011hw is located at $RA = 22^h26^m14^s.54, Dec = +34^o12'59''.1$ (equinox J2000.0), approximately 8 arcsec East and 1 arcsec North of the center of an
anonymous host galaxy \citep{din11}, which is not listed in the major galaxy catalogues (Figure \ref{fig:app_maps}, bottom). 
The only way to estimate its distance is through the redshift as deduced from the shift of the SN spectral features. \citet{val11} determined a redshift of $z$ = 0.023 $\pm$ 0.001; a similar value was estimated 
by \citet{smi12}. Assuming H$_0$ = 73 km s$^{-1}$ Mpc$^{-1}$ ($\Omega_M$ = 0.27 and $\Omega_\Lambda$ = 0.73), 
we obtain a luminosity distance of 96.2 Mpc, which implies $\mu$ = 34.92 $\pm$ 0.24 mag.
\citet{sch11} estimate a Milky Way contribution to the extinction  of 
 $A_B$(MW) = 0.42 mag. Since available spectra do not show any evident narrow interstellar line at the host galaxy rest frame, 
we assume that there is no host galaxy contribution to the total reddening. Therefore $A_B$(tot) =  $A_B$(MW) = 0.42 mag.

\subsection{Light Curves}  \label{light_curves}

\begin{figure*}
   \centering
  \includegraphics[angle=270,width=7.15in]{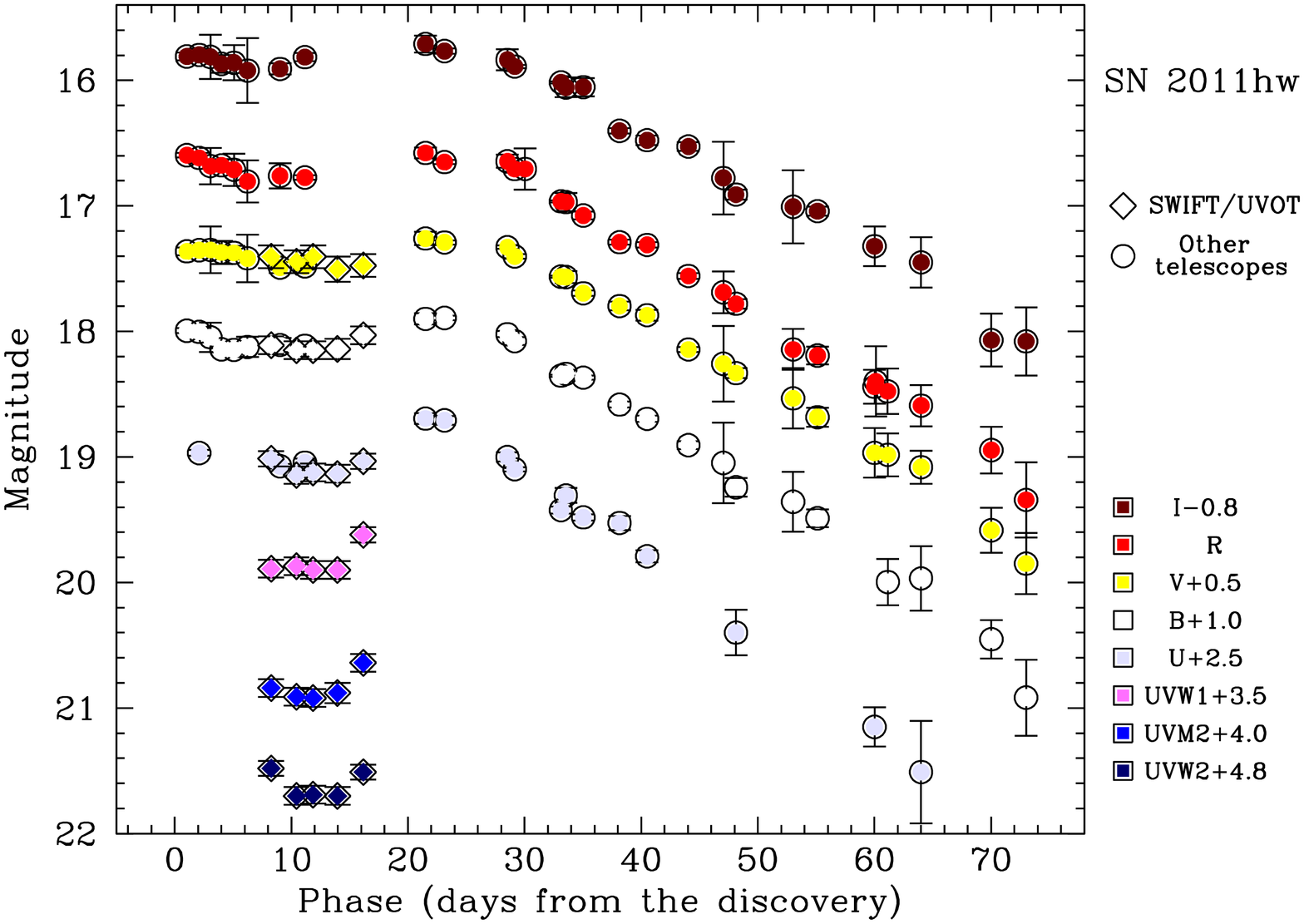} 
   \caption{
UV and optical light curves of SN 2011hw. Swift/UVOT optical-band magnitudes have been shifted by $+$0.054 mag in $U$ and $-$0.025 mag in $V$
to match the ground-based measurements.  }
   \label{fig:11hw_LC}
\end{figure*}

The two SNe have different pre-discovery histories. SN 2010al was discovered on 2010 Mar 13.03 UT ($JD$ = 2455268.53) 
and nothing was visible at the SN position in archive images obtained on 2010 Feb 7.12 UT \citep[$JD$ = 2455234.62,][]{rich10}. This detection limit alone cannot constrain well the
explosion epoch. However, pre-discovery images obtained on Mar 12.71 UT ($JD$ = 2455268.21) during routine observations performed with
the 0.4-m MASTER telescope at Kislovodsk (Caucasian region, Russia) do not show any evidence of the SN to a limiting magnitude of
$R$ = 19.2. The fast rise of the light curve (see below) and the early spectra (Section \ref{spec}) confirm that the object was discovered very young, close to 
the core-collapse epoch. Hereafter, we will adopt $JD$ = 2455268.0 $\pm$ 1.5 as the time of the explosion.

On the contrary, the explosion epoch of SN 2011hw is not well known. \citet{din11} discovered SN 2011hw on 2011 Nov 18.72 UT ($JD$ = 2455884.22), 
at an apparent unfiltered magnitude of 15.7.  Our earliest follow-up observation was obtained one day after the discovery, 
and the SN magnitude was  estimated to be R $\approx$ 16.6, i.e. almost 1 mag fainter than the magnitude reported by \citet{din11}. 
We note that our photometry data are in decent agreement (within a few tenths mag) with those of  \citet{smi12}.
The last prediscovery image with negative detection was on 2010 Dec 12 (with a limiting magnitude of 19.5), almost one year 
before \citep{din11} and then does not tightly constrain the explosion epoch. However, a comparison of the first spectrum of SN 2011hw with the spectra of other Type Ibn SNe suggests that SN 2011hw
was discovered quite late (see Section \ref{spec}), at least a couple of weeks after the core-collapse. We adopt
November 4th, 2011 ($JD$ = 2455870 $\pm$ 10) as the epoch of the explosion.

\begin{figure*}
 {\includegraphics[width=3.46in]{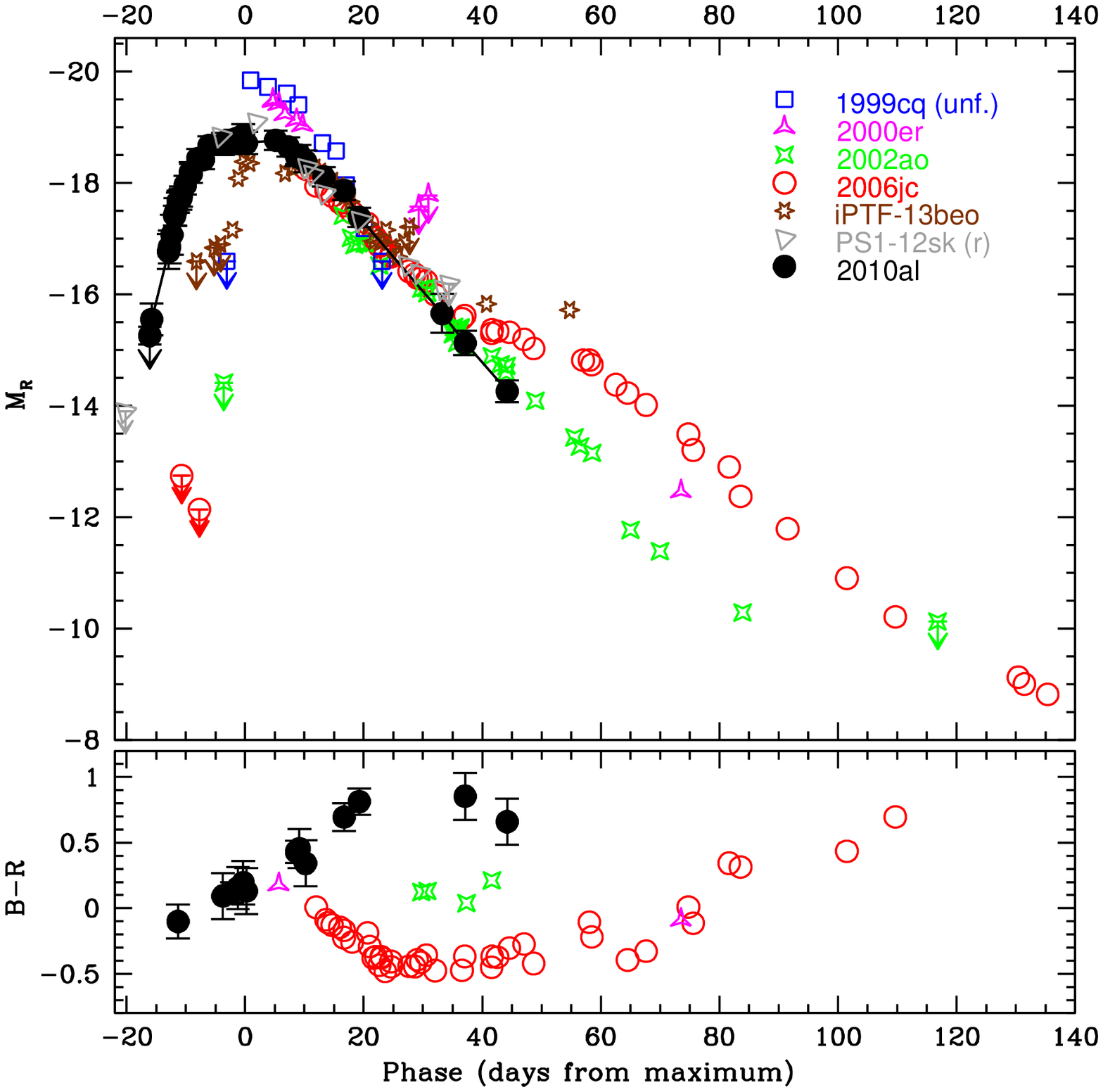}
  \includegraphics[width=3.46in]{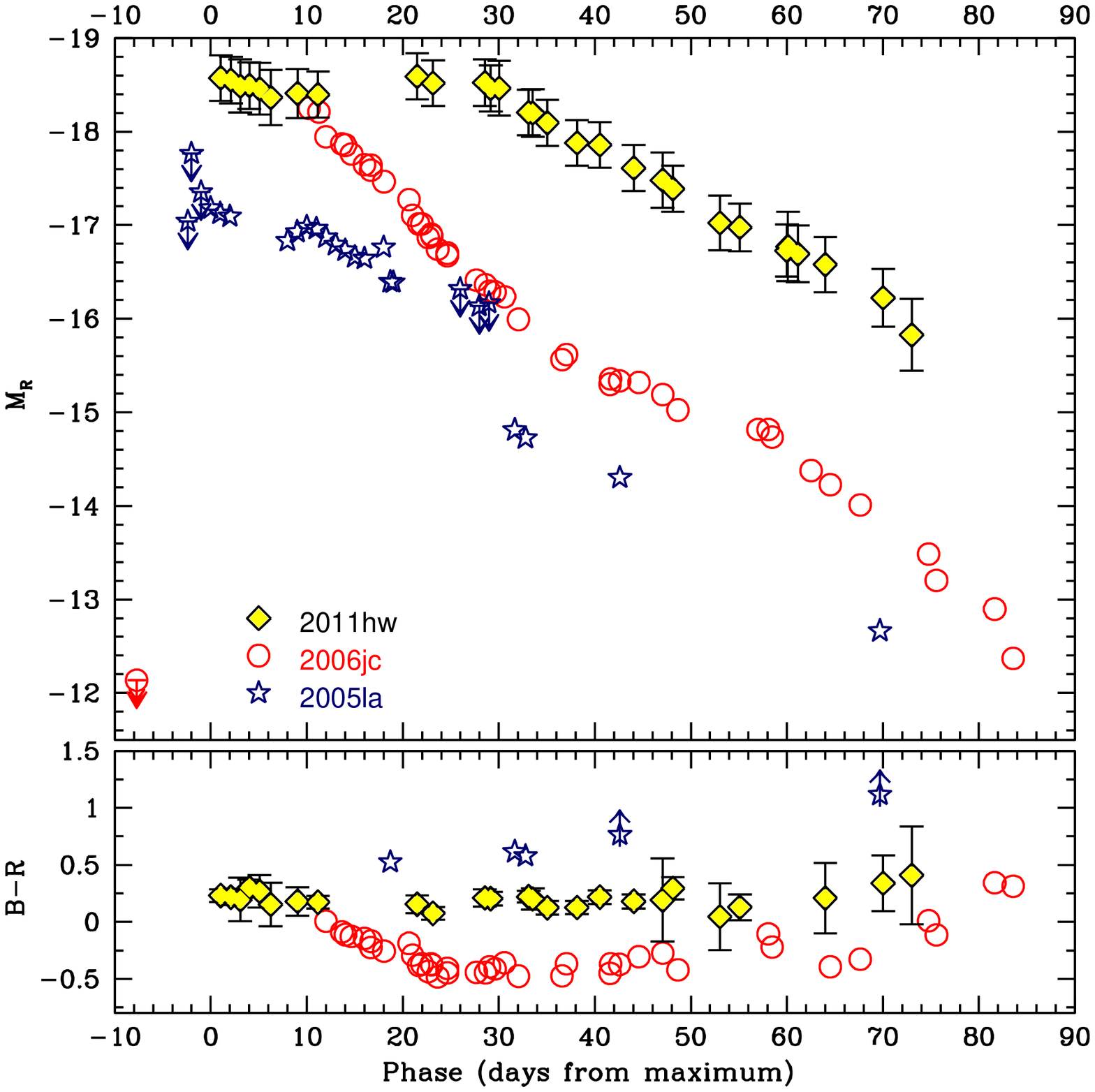}}
\caption{{\bf Top-left:} $R$-band absolute light curves of SN 2010al and other Type Ibn SNe, including SNe 1999cq, 2000er, 2002ao, 2006jc, PS1-12sk and iPTF-13beo  \protect\citep{mat00,pasto07,pasto08a,fol07,san13,gor14}. SN 2010al is marginally
fainter at peak than other SNe Ibn shown here, and has a slower rise to maximum. {\bf Bottom-left:} $B-R$ colour curves for the same sample of SNe Ibn.
{\bf Top-right:} $R$-band absolute light curves of the Type Ibn/IIn SNe 2011hw and 2005la, compared with the prototypical SN Ibn 2006jc. For SNe 2011hw and 2005la, we 
adopt the discovery epochs as reference dates for the maximum. For SN 2005la, only the most significant detection limits are shown. {\bf Bottom-right:} $B-R$ colour curves for SNe 2011hw, 2005la and 2006jc.
Data have been corrected for interstellar reddening.
\label{fig_abs}}
\end{figure*}

SN 2010al has been extensively targeted by SWIFT + UVOT, from the UV domain to the blue optical bands. 
Unfortunately, the optical and NIR ground-based coverage is not ideal, suffering from some observational gaps.
However, additional contributions of unfiltered photometry from amateur astronomers and $R$-band photometry obtained with MASTER facilities allowed us to
to improve the light curve sampling at least in the $R$ band. 
Additional multi-band photometry measurements as derived from the accurately calibrated XShooter spectra \footnote{R-band VLT acquisition images were used to properly scale in flux
the XShooter spectra, and spectro-photometric measurements were performed using the package {\sl STSDAS} in IRAF. A final uncertainty of 10 per cent in the flux calibration has
been adopted for the XShooter spectra (Section \ref{spec}).} were also used  (see Table \ref{tab_ph_10al}).
The resulting  light curves of SN 2010al are shown in Figure \ref{fig:10al_LC}. The left panel of Figure \ref{fig:10al_LC} shows the early UV-to-NIR evolution (up to $\sim$2 months), 
whilst the right panel includes also the late-time optical+NIR observations obtained when the SN was recovered after the seasonal observational gap. 
At late times the object was detected only once in the NIR bands, while
it was not detectable in optical observations, and we could only estimate upper detection limits. We note that the decline rate in the NIR bands after the seasonal gap is marginally flatter 
than the decline rate expected when the $^{56}$Co decay to $^{56}$Fe powers the SN light curve (assuming full gamma-ray trapping). 
Unfortunately, no information on the decline rate can be inferred from the late-time optical observations.
 
\begin{figure}
   \centering
  {\includegraphics[width=3.45in]{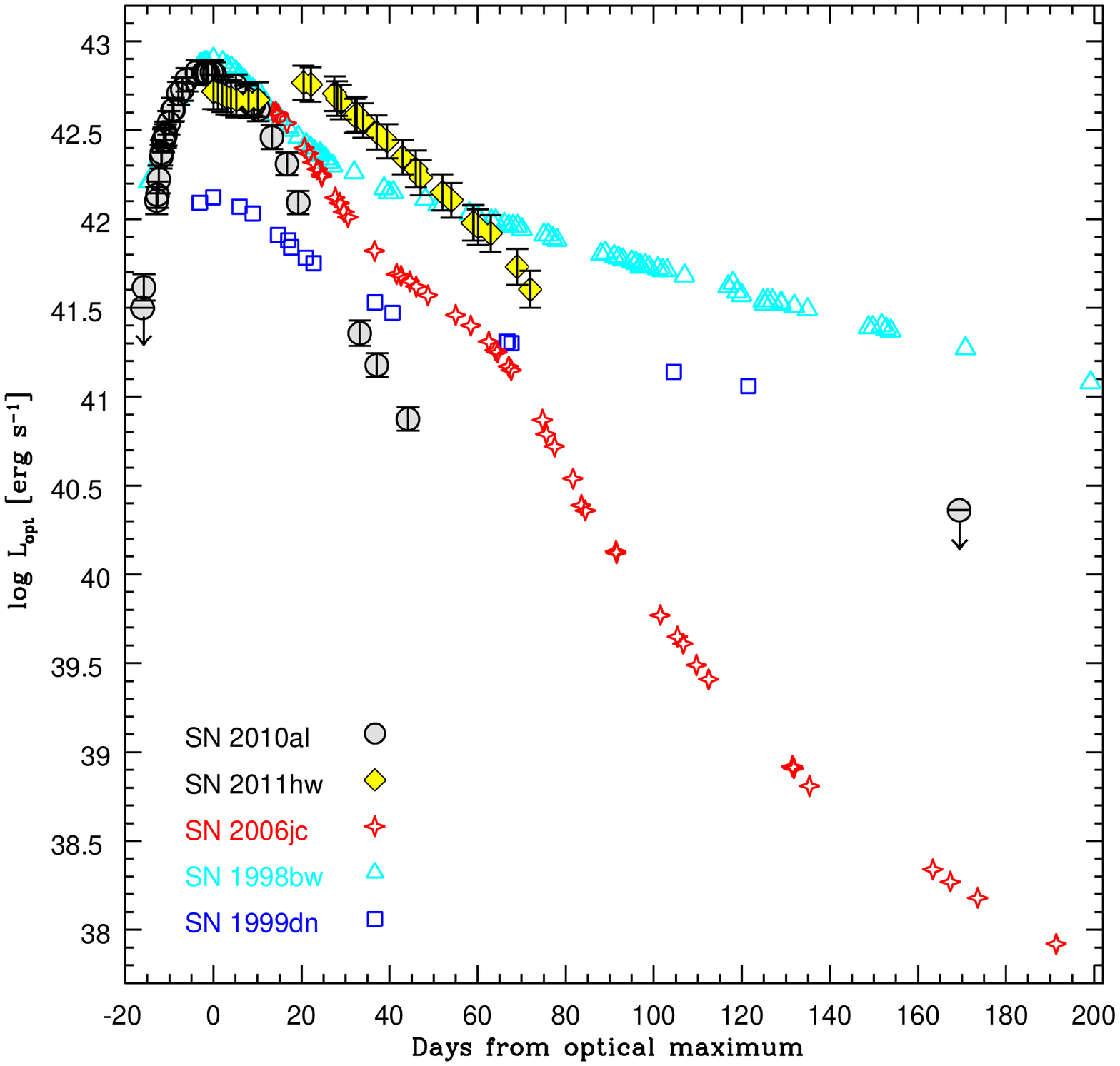} \includegraphics[width=3.45in]{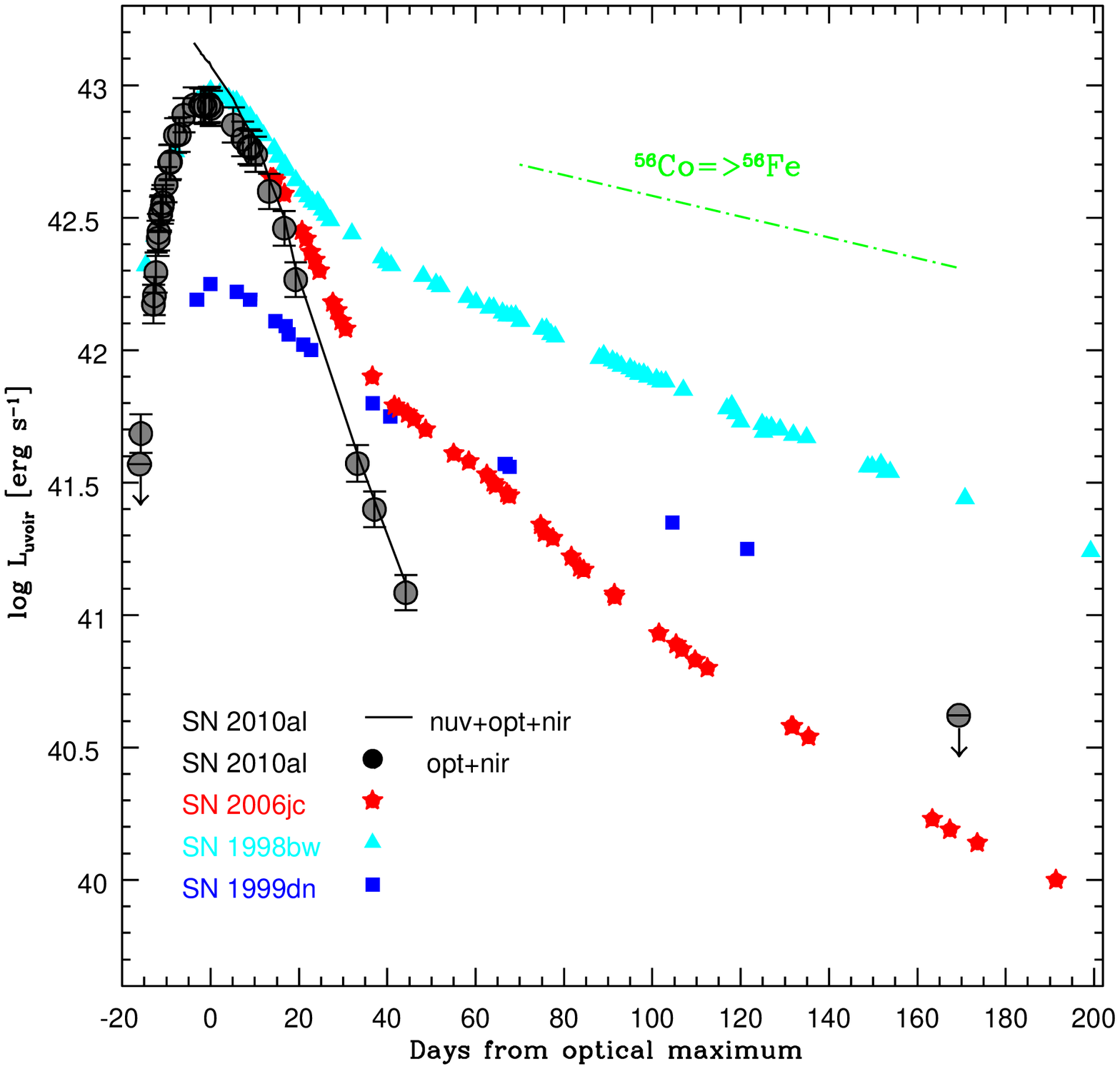}}
   \caption{Quasi-bolometric light curves of the Type Ibn SNe 2006jc, 2011hw and 2010al compared with those of the normal SN Ib 1999dn \citep{ben11} and the broad-lined Ic SN 1998bw \citep[][and references therein]{pat01}. {\bf Top:} Pseudo-bolometric light curves for the SN sample have been computed by integrating the fluxes in the optical bands only. 
{\bf Bottom:} The pseudo-bolometric light curves have been obtained by integrating the broadband fluxes in the optical and NIR domains. In the case of SN 2010al, also an {\it uvoir} curve obtained 
including the near-UV contribution is shown (solid line), although limited to phases later than -3.7 days from maximum. Error-bars have been reported for SNe 2010al and 2011hw, and account for 
uncertainties in the distance and the reddening estimates, spectral energy distribution fitting and photometric errors.}
   \label{bolo_LC}
\end{figure}

The photometric follow-up campaign of SN 2011hw in the optical bands lasted about 2.5 months, after which the object disappeared behind the sun. Additional late-time imaging was obtained 
about 6-7 months later, but the object was below the detection threshold. In fact, no source was detected at the SN position in TNG+Dolores images obtained on July 25th, 2012 (to limiting magnitudes
$R$ = 23.2 and $I$ = 23.1) and in very deep WHT+ACAM images obtained on August 21st, 2012 (to limiting magnitudes $U$ = 23.3, $B$ = 23.9, $V$ = 24.2).
The light curve (see Figure \ref{fig:11hw_LC}) is peculiar, showing a modest decline in all bands soon after the discovery
lasting about 1 week. This was followed by a re-brightening leading to a second maximum at phase of about 3 weeks. This was already noted by \citet{smi12}, 
and interpreted as an additional luminosity input from interaction when the shock reaches a higher density CSM shell. Similar re-brightenings in the light curves
have been onserved also in the transitional Type Ibn/IIn SN 2005la \citep{pasto08b} and the Type Ibn iPTF13beo \citep{gor14}.

The secondary maximum is then followed by a fast linear decline, which is steeper in the
blue bands, with slopes: $\Delta U$ = 6.8 $\pm$ 0.2 mag/100$^d$; $\Delta B$ = 5.8 $\pm$ 0.2 mag/100$^d$; $\Delta V$ = 5.2 $\pm$ 0.2 mag/100$^d$;  
$\Delta R$ = 5.5 $\pm$ 0.1 mag/100$^d$; $\Delta I$ = 4.7 $\pm$ 0.2 mag/100$^d$.
Additional early-time UV photometry was obtained with  SWIFT/UVOT showing the same  
re-brightening to the secondary maximum as the optical bands. Unfortunately, the SWIFT campaign was suspended after 8 days, and no UV observations were obtained during the second peak.

In Figure \ref{fig_abs} (top-left panel) the absolute $R$-band light curve of SN 2010al is shown along with those of the Type Ibn SNe 1999cq \citep{mat00}, 2006jc \citep{pasto07,pasto08a,fol07},  
 PS1-12sk \citep{san13}, iPTF-13beo \citep{gor14}, 2000er and 2002ao \citep{pasto08a}.
SN 2010al (peaking at an absolute magnitude $M_R$ = $-$18.86 $\pm$ 0.21 on $JD$ = 2455284.3 $\pm$ 1.1) is slightly less luminous than both SN 2000er and SN 1999cq, and has a symmetric light curve peak, with a slow rise
but a similar post-peak decline rate. In the top-right panel of 
Figure \ref{fig_abs}, the light curve of SN 2011hw is compared with those of SN 2006jc and the Type Ibn/IIn 
SN 2005la \citep{pasto08b}. SN 2011hw has an absolute magnitude at the second maximum of $M_R$ = $-$18.59 $\pm$ 0.25, which is very close to the magnitude at discovery.
Although the luminosity of SN 2005la is lower and its light curve has some scatter, the indication is that SN 2005la,
like SN 2011hw, has a non-monotonic behaviour after the first maximum, possibly related to an enhanced contribution of the ejecta interaction with a higher-density CSM. A shoulder is also visible in the optical light curve of  SN 2006jc
after $\sim$ 40 days, but it was shown to be related to the formation of dust in a post-shock cool dense shell \citep[see e.g.][]{smi08,seppo08}. One of the effects of dust formation is the attenuation of the
light at optical wavelengths and the  enhanced emission in the infrared domain. The lack of NIR observations does not allow us to verify whether dust has formed in the ejecta of SN 2011hw or in the circumstellar environment.

In the bottom panels of Figure \ref{fig_abs} the extinction-corrected $B-R$ colour curves of our Type Ibn SN sample are shown. The comparisons show that there is some heterogeneity in the colour
evolution of SNe Ibn. In comparison with the other objects, SN 2010al shows an opposite colour trend (bottom-left panel), reaching a $B-R$ colour maximum of $\sim$1 mag at about 30 days past maximum, while at the same phase the 
colour of SN 2006jc was extremely blue ($B-R$ $\approx$ $-$0.5 mag). 
After $\sim$40 days from maximum, SN 2006jc and the transitional SNe 2011hw and 2005la show a moderate trend toward  redder colours  
(bottom-right panel), which may be a signature of dust formation. However, we agree with the findings of \citet{smi12} that the available spectra of SN 2011hw do not show a clear evidence 
for blue-shifted spectral line peaks, and this would argue against the dust formation in this object.

\subsection{Quasi-bolometric light curves} \label{bolo}

Quasi-bolometric light curves for SNe 2010al and 2011hw have been computed either by integrating the fluxes in the optical bands only, and - in the case of SN 2010al - including also the UV and 
NIR contributions to the total 
flux. Occasionally, photometric data in a given filter were not available. The flux contribution of the missing band was then obtained by interpolating
the fluxes between epochs when photometric observations were available or, when necessary, by extrapolating the missing photometry from the earliest/latest available epoch, assuming  a constant colour evolution.
The flux in individual bands was corrected for the adopted extinction and used to derive the spectral energy distribution that was then integrated to derive the quasi-bolometric flux.
The total flux was finally converted in luminosity using the extinction values and the distance moduli discussed in Section \ref{red_dis}.
The resulting quasi-bolometric light curves are shown in Figure \ref{bolo_LC}, along with those of the prototypical SN Ibn 2006jc \citep{fol07,pasto07,pasto08a,seppo08}, 
the normal SN Ib 1999dn  \citep{ben11} and the broad-line SN Ic 1998bw \citep{gal98,mck99,pat01,sol02}. The top panel of Figure  \ref{bolo_LC} shows the ``optical'' pseudo-bolometric light curves,
the bottom panel reports the pseudo-bolometric light curves obtained by including also the flux contribution in the NIR domain. 
For SN 2010al, the UV to NIR ({\it uvoir}) curve is also shown, plotted with a solid line. 
We note that we can provide only a lower limit (L $>$ 5.5 $\times$ 10$^{42}$ erg s$^{-1}$) 
for the maximum luminosity of SN 2011hw, since the object was probably discovered after maximum, and NIR observations were not available.

The quasi-bolometric light curve of SN 2010al peaks at a maximum luminosity of about 10$^{43}$ erg s$^{-1}$, which is similar to the peak luminosity of SN 1998bw. We also note that 
 NIR contribution is small around maximum, but increases with time becoming significantly larger at post-peak phases.  
A similar behaviour was also observed in SN 2006jc, although in that case the late NIR contribution was much more significant. A strong deficit in the optical light curve was observed at $>$40 days past maximum 
\citep[][see also Figure \ref{bolo_LC}]{smi08,seppo08} and was balanced by a clear NIR excess. The resulting quasi-bolometric light curve of SN 2006jc showed a decline that was 
consistent with a $^{56}$Co-powered event assuming a complete $\gamma$-ray trapping (Figure \ref{bolo_LC}, bottom). 
We remark that in SN 2010al there is a single late-time detection of the SN in the NIR bands and no detection in the optical bands at nebular phases. 
So, the luminosity estimate at $\sim$162 days in Figure \ref{bolo_LC} has to be regarded as an upper limit and, hence,
 we cannot provide a reliable estimate for the $^{56}$Ni mass.

\begin{table*}
\caption{Log of spectroscopic observations of SNe 2010al and 2011hw.}
\begin{center}
\footnotesize
\begin{tabular}{ccccccc} \hline\hline
\multicolumn{6}{c}{SN 2010al} \\ \hline
Data & JD+ & \multicolumn{2}{c}{Phase (days)} & Instrumental configuration & Range (\AA) & Resolution$^\ast$ (\AA)  \\ 
   & 2455000 & after core-collapse & after discovery & & & \\ \hline
20Mar10 &  275.65 & 7.7 & 7.1 & DuPont + B$\&$C & 3700-9250 & 7 \\ 
25Mar10 &  280.55 & 12.6 & 12.0 & VLT-UT2 + XShooter & 3000-24800 & 0.8;0.8;3.2 \\
29Mar10 &  284.50 & 16.5 & 16.0 & VLT-UT2 + XShooter & 3000-24800 & 0.8;0.8;3.2 \\
30Mar10 &  286.38 & 18.4 & 17.9 & WHT + ISIS & 3000-10000 & 4.9;9.5 \\
01Apr10 &  288.47 & 20.5 & 19.9 & NOT + ALFOSC & 3200-9000 & 18 \\
06Apr10$^\ddag$ &  293.42 & 25.4 & 24.9 & Ekar182 + AFOSC & 4100-8100 & 24 \\
07Apr10 & 294.49 & 26.5 & 26.0 & VLT-UT2 + XShooter & 3000-24800 & 0.8;0.8;3.2 \\
18Apr10 & 304.58 & 36.6 & 36.1 & NTT + EFOSC2 & 3650-9050 & 27 \\
19Apr10 & 305.56 & 37.6 & 37.0 & NTT + SOFI & 9350-16450 & 25 \\
22Apr10 & 309.42 & 41.4 & 40.9 & TNG + NICS & 8700-14550 & 19 \\
24Apr10 & 311.47 & 43.5 & 42.9 & TNG + LRS & 3200-7950 & 15 \\
29Apr10 & 316.44 & 48.4 & 47.9 & TNG + LRS & 5050-9350 & 14 \\
05May10 & 322.42 & 54.4 & 53.9 & TNG + LRS & 5000-10200 & 9.5 \\
11May10 & 328.48 & 60.5 & 60.0 & VLT-UT2 + XShooter & 3000-24800 & 0.8;0.8;3.2 \\ 
\hline\hline
\multicolumn{6}{c}{SN 2011hw} \\ \hline
Data & JD+ &  \multicolumn{2}{c}{Phase (days)} & Instrumental configuration & Range (\AA) & Resolution$^\ast$ (\AA)  \\
   & 2455000 & after core-collapse & after discovery & & & \\ \hline
19Nov11 & 885.26 & 15.3 & 1.0 & Ekar182 + AFOSC & 3550-8200 & 12;24 \\
20Nov11 & 886.30 & 16.3 & 2.1 & Ekar182 + AFOSC & 3500-8150 & 24 \\
24Nov11 & 890.43 & 20.4 & 6.2  & Ekar182 + AFOSC & 3600-8200 & 24 \\
27Nov11 & 893.32 & 23.3 & 9.1 & Ekar182 + AFOSC & 3600-8200 & 12   \\
28Nov11 & 894.40 & 24.4 & 10.2 & NOT + ALFOSC & 3300-9100 & 14 \\
29Nov11 & 895.43 & 25.4 & 11.2 & CAHA2.2 + CAFOS & 5800-9600 & 6 \\ 
17Dec11 & 913.34 & 43.3 & 29.1 & TNG + LRS & 3250-10350 & 15;14\\
18Dec11$^\ddag$ & 914.29 & 44.3 & 30.1 & Ekar182 + AFOSC & 3500-8200 & 24 \\ 
21Dec11 & 917.31 & 47.3 & 33.1 & WHT + ISIS & 3100-10400 & 5;9 \\
01Jan12 & 928.31 & 58.3 & 44.1 & CAHA2.2+CAFOS & 3350-8850 & 14 \\
17Jan12 & 944.36 & 74.4 & 60.1 & TNG + LRS & 3300-8050 & 15 \\
18Jan12 & 945.36 & 75.4 & 61.1 & TNG + LRS & 5050-9650 & 14 \\
29Jan12 & 956.36 & 86.4 & 72.1 & NOT + ALFOSC & 3400-9100 & 18  \\
\hline
\end{tabular}

$^\ast$ As measured from the FWHM of the night sky lines.
$^\ddag$ Poorer signal-to-noise spectra, not shown in Figures \ref{sn10al_spec} and \ref{sn11hw_spec}.

\end{center}
\label{tab_log_spec}
\end{table*}

\section{Spectroscopy} \label{spec}

Both SN 2010al and SN 2011hw have extensive spectroscopic sequences, and the reduction of the spectra
was performed using standard {\it IRAF} tasks. The preliminary reduction steps included overscan and bias corrections, flat-fielding and trimming, following
the same prescriptions as imaging data. For the NIR spectra, the contribution of the night sky background was removed by subtracting from each other two consecutive exposures 
taken with the source in different positions along the slit. 
The spectra were then optimally extracted to remove all background contamination and hot pixels. Then, spectra of the science targets were wavelength calibrated using arc lamp comparison spectra,
and were finally flux-calibrated using sensitivity curves obtained from spectro-photometric standard star spectra. 

The higher resolution XShooter spectra were processed using of the dedicated ESO pipeline\footnote{\it http://www.eso.org/sci/software/pipelines/}.
For each of the UV, optical and NIR channels, linearized, sky-subtracted and wavelength-calibrated 2-dimensional spectra were obtained from  the curved Echelle 
orders of XShooter. The 1-dimensional spectra were obtained through the optimal extraction as for the low-resolution spectra.
Relative flux calibrations were performed through spectro-photometric standards for which flux tables extending from the UV to the NIR domains are available.
The flux tables were taken from the dedicated ESO web site.\footnote{\it https://www.eso.org/sci/observing/tools/standards/spectra.html}

For all spectra, the accuracy of the 
spectroscopic flux calibration was checked using the available SN photometry. In case of discrepancy, the spectral fluxes were rescaled to
match the photometric data. The expected uncertainty in the flux calibration is about 10 per cent.
Finally, spectra of telluric standards were used to remove the broad atmospheric absorption bands from the SN spectra.

The log containing the spectroscopic observations of the two events is provided in Table \ref{tab_log_spec}.

\subsection{Spectral sequence of SN 2010al} \label{sec:spec10al}

\begin{figure*}
%   \centering
  { \includegraphics[angle=270,width=6.0in]{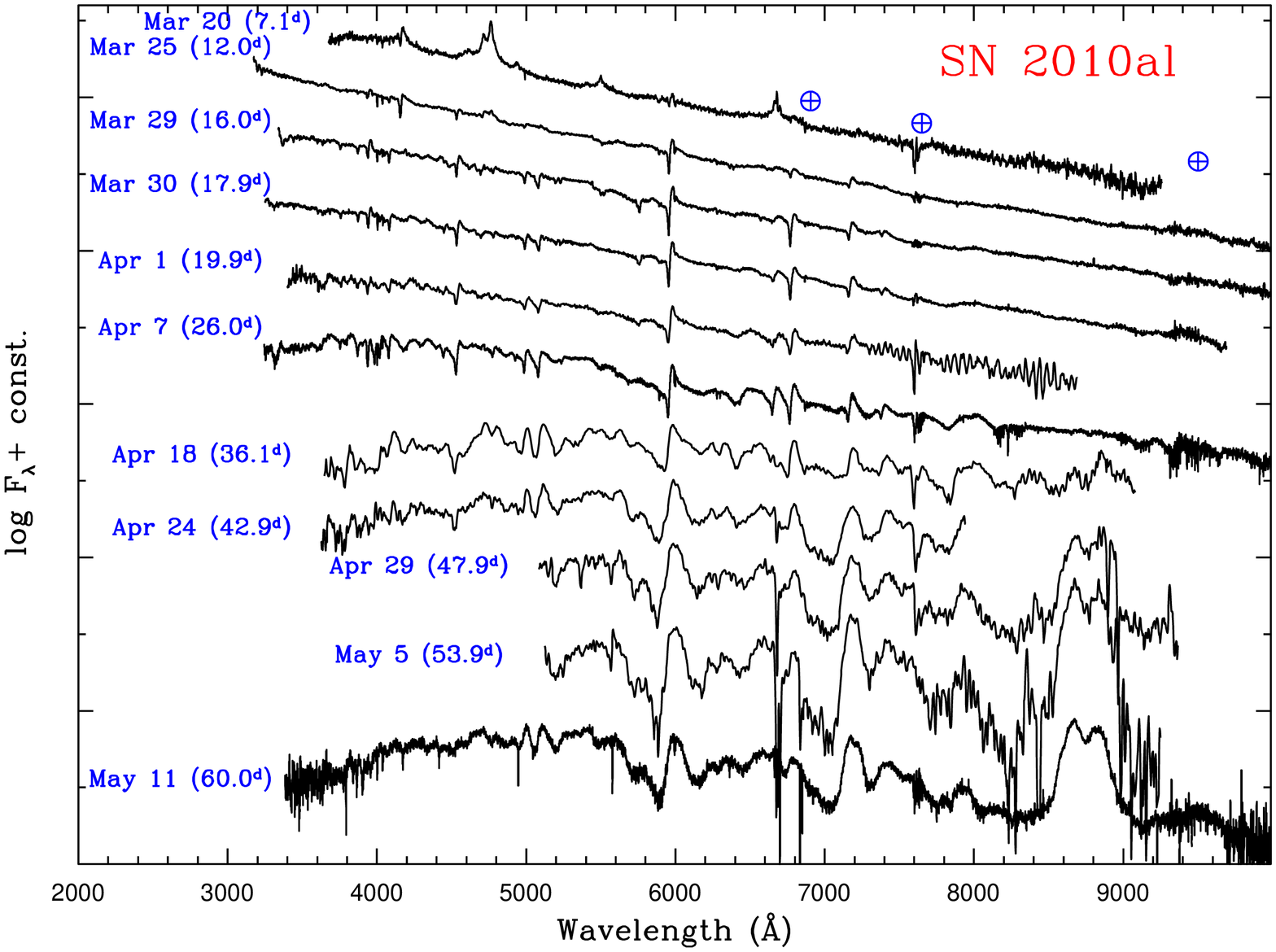} 
   \includegraphics[angle=270,width=6.0in]{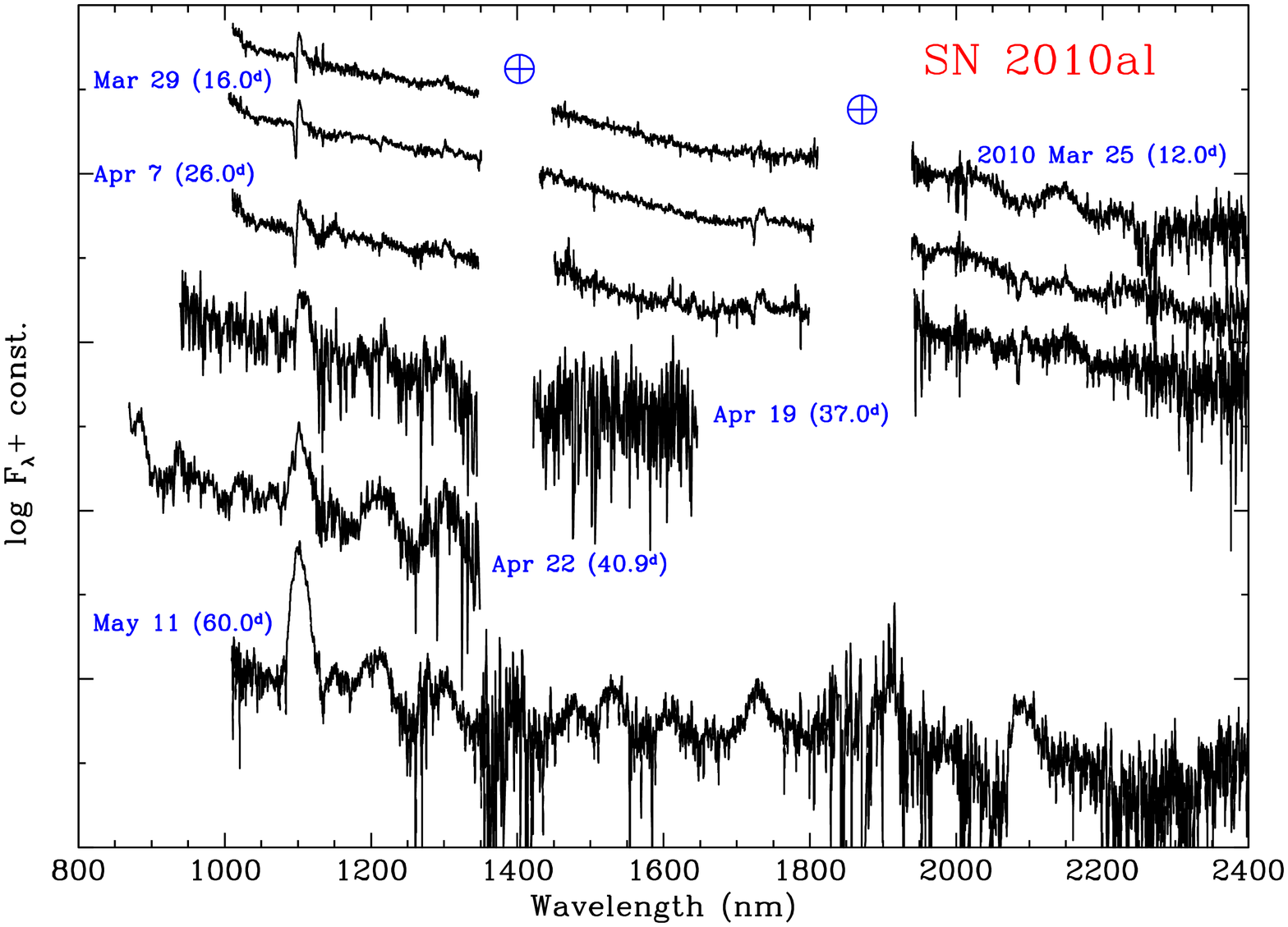} }
   \caption{Optical ({\bf top panel}) and NIR ({\bf bottom panel}) spectra of SN 2010al. No redshift or reddening corrections have been applied. The positions of the most important
telluric bands are marked with ``$\oplus$''. The phases reported in brackets are days after discovery.}
   \label{sn10al_spec}
\end{figure*}

SN 2010al was observed in optical and NIR spectroscopy from day 7 after the discovery to about 
day 60.  In Figure \ref{sn10al_spec} the sequence of our good signal-to-noise spectra of 
SN 2010al is shown, with the optical spectra being in the top panel, and the NIR spectra in the bottom panel. 

Our earliest spectrum shown in Figure  \ref{sn10al_spec} is the classification spectrum of \citet{max10}. The spectrum is peculiar, showing a blue continuum and relatively prominent Balmer lines in emission. 
H$\alpha$ has a narrow unresolved component ($<$ 360 km s$^{-1}$) possibly due to interstellar gas contamination, superposed on an intermediate component with a full width at half maximum (FWHM) 
velocity $v_{FWHM} \approx$ 1800 km s$^{-1}$.

\begin{figure}
   \centering
  \includegraphics[width=3.45in]{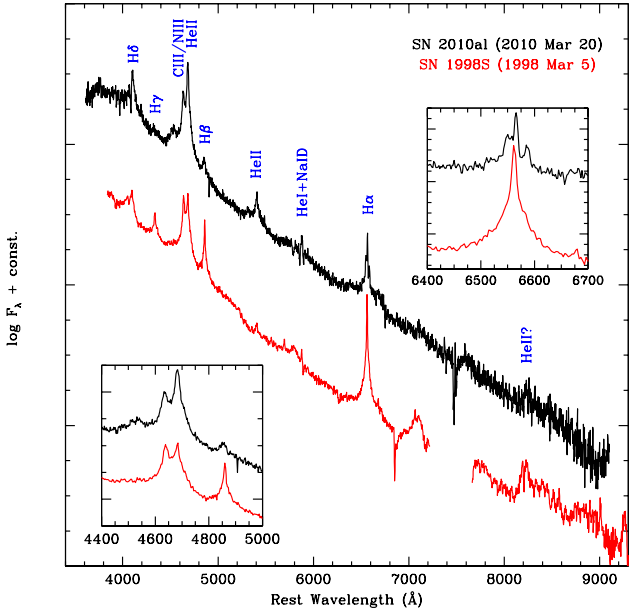} 
   \caption{Comparison between the earliest optical spectra of SN 2010al and the Type IIn SN 1998S \protect\citep{fas01}. The two spectra are reddening- and redshift-corrected. The inserts shows a blow up of the region between 4400 and 5000 \AA~(bottom-left) and the region of H$\alpha$ (top-right) in the two spectra.}
   \label{fig_98S}
\end{figure}

The most prominent emission feature lies in the blue region of the spectrum (at about 4660 \AA) and shows a double-peaked profile. The redder component peaks at 4684 \AA~ and is very likely He II $\lambda$4686, while the bluer
emission peaks at 4646 \AA, and is probably a blend of  C III $\lambda$4648 and N III $\lambda$4640. We note that these lines, identified also by \citet{coo10} and \citet{sil10}, are often observed in Wolf-Rayet winds. 
A very similar feature was also observed in very early spectra of the Type IIn SNe 1998S \citep{fas01}, 2008fq \citep{tad13}, and also in the classification spectrum of LSQ13fn \citep{sol13}. 
A comparison with the spectrum of SN 1998S at 3.3 days after the discovery presented by \citet{fas01}
is shown in Figure \ref{fig_98S}; the two inserts emphasize the regions of the peculiar spectral feature at $\sim$ 4600-4700 \AA~ and 
that of H$\alpha$.
The identification of He II in both SNe is safe, since He II $\lambda$5411 is also detected, with $v_{FWHM} \approx$ 2250 km s$^{-1}$. 
Another, weaker feature is detectable at about 4460-4560 \AA~ and is probably a line blend of various ionized elements, including He II $\lambda$4541, although we cannot exclude a minor contribution from the 
He I $\lambda$4471 line (see the insert in Figure \ref{fig_98S}). We also tentatively identify He II $\lambda$8236, following \citet{fas01}.

The second spectrum of SN 2010al (day 12 after the discovery) is very different, and this suggests us to revise the classification of this object as a Type Ibn SN (see Section \ref{intro}). 
The spectrum is still dominated by a blue continuum ($T_{bb}$ = 12800 $\pm$ 400 K), but now  
the most remarkable lines visible in the spectrum are He I, with unusually narrow P-Cygni profiles.
 The position of the minimum of the blue-shifted absorption component suggests velocities of the He-rich material of about 1000-1100 km s$^{-1}$. The feature detected in the first
spectrum at about 4600-4700 \AA~is now much weaker, although it still shows a double-peaked profile. 
A very weak hump probably due to H$\alpha$ is also detected.
Very narrow absorptions of Ca II H$\&$K and Na ID are attributed to material lying along the line of sight and are unrelated to the
SN environment.

In the third spectrum (day 16, $T_{bb}$ = 11900 $\pm$ 400 K), He I  P-Cygni lines with a measured expansion velocity $v \approx$ 1050-1150 km s$^{-1}$
become more and more prominent. We note that these P-Cygni He I lines are likely produced in He-rich CSM moving at a velocity of above 1000 km s$^{-1}$ 
which was initially ionized (e.g. at the epoch of our first spectrum), and is now recombining.
At this phase, together with the He I lines, probably other features (including weak Fe II lines) are detected.
The Wolf-Rayet feature at about 4600-4700 \AA~ has  completely  disappeared, while H$\alpha$ is still barely detected (with expansion velocity of about 1100 km s$^{-1}$).
The following spectra (phases 18 and 20 days after discovery) do not show a significant evolution.
  
\begin{figure*}
\begin{center}
   \includegraphics[width=6.4in]{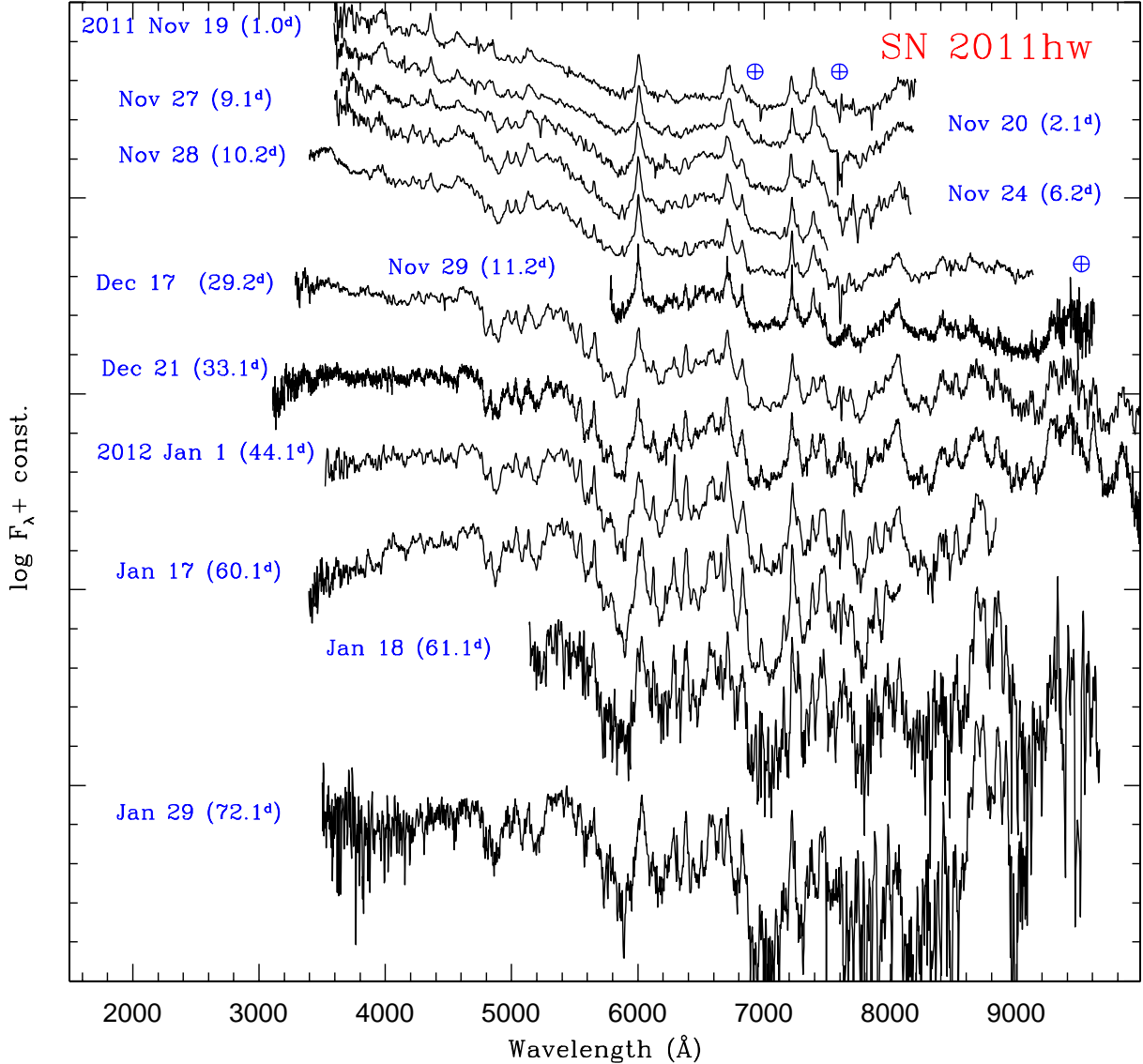} 
\caption{Optical spectral sequence for SN 2011hw. No redshift or reddening corrections have been applied. The positions of the most important
telluric bands are marked with ``$\oplus$''. The phases reported in brackets are days after discovery.}
\label{sn11hw_spec}
\end{center}
\end{figure*}

\begin{figure*}
\begin{center}
   \includegraphics[angle=270,width=7.0in]{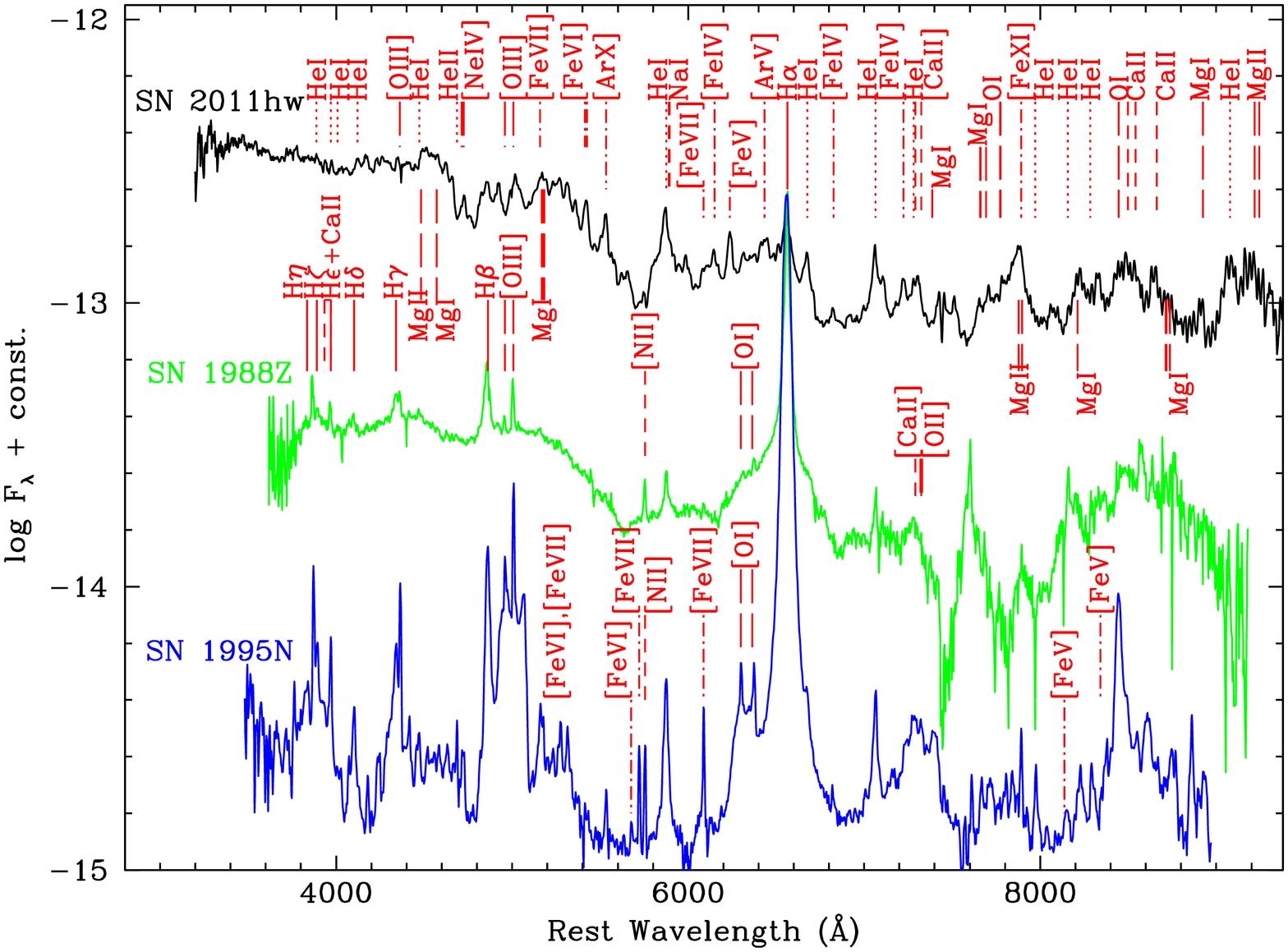} 
\caption{The 2011 December 17 spectrum of SN 2011hw is shown together with spectra of the Type IIn SNe 1988Z and 1995N. The most important lines detectable in the spectra are marked. }
\label{lineID11hw}
\end{center}
\end{figure*}

The XShooter spectrum at day 26  has a redder continuum ($T_{bb}$ = 9100 $\pm$ 500 K), and  many new P-Cygni SN lines are now detected, including Ca II H$\&$K and  Fe II.
A few weak lines of He I are now visible in the NIR region. In particular, the He I $\lambda$20581 line is clearly detected.
In addition, H$\alpha$ is visible with a P-Cygni profile, and with an expansion velocity that is comparable with that of the
He I, Ca II and Fe II lines, i.e. around 1300-1400 km s$^{-1}$.

The spectrum obtained 36 days after discovery shows major changes. The continuum is now much redder and (especially at red wavelengths)
the emission components start to dominate over the absorptions. In addition the lines are broader (1900-2300 km s$^{-1}$). H$\alpha$ is not visible anymore, and the NIR Ca II
triplet is now clearly detected. A broad line likely due to a blend including also O I $\lambda$$\lambda$7772-7775 is one of the most prominent spectral features. The two NIR spectra at days 37 and 41,
despite the low signal-to-noise, show that the most prominent line in the NIR region is He I $\lambda$10830, almost purely in emission, with a FWHM velocity of about 5000 km s$^{-1}$.

The following spectra (days 43, 48 and 54) show a strong pseudo-continuum  below 5600 \AA~(which is much stronger in SN 2011hw, see below), 
in analogy with that  observed in other SN 2006jc-like events (but also in Type IIn SNe). The lines have a similar width as 
in the previous NIR spectra (4000-5000 km s$^{-1}$), and the most prominent feature is now the broad  NIR Ca II triplet in emission, with a double-peaked profile and a total 
FWHM of about 12000 km s$^{-1}$.
The last XShooter spectrum (day 60) confirms most characteristics of the previous spectra, with a significant blue pseudo-continuum where He I and Fe II absorptions are still well
visible. The lines of He I $\lambda$5876 (possibly blended with  Na ID), $\lambda$7065 and $\lambda$10830 are now prominent in emission although a residual broad P-Cygni absorption is still barely
visible. Other He I lines are detected in the NIR, in particular $\lambda$15084, $\lambda$17002, $\lambda\lambda$18685-18697 and $\lambda$20581.
The FWHM velocity of all these lines from a Gaussian fit is about 5000 $\pm$ 1000 km s$^{-1}$. The apparent broadening of the spectral lines with time indicates
that the photosphere recedes with time from the shocked CSM to the SN ejecta. 
This explanation is supported by the Gaussian emission lines 
that do not show evident boxy profiles. Other features that can be observed in the day 60 spectrum are a very prominent Ca~II NIR triplet in emission
(blended with O I $\lambda$8448) and Mg I $\lambda$7659.
Mg lines can also contribute to shaping the apparent emission bumps at $\sim$ 4600 \AA, 5200 \AA~and 9000-9500 \AA~(blended with O I).
The broad bump around 6600 \AA~can be due to a blend of different lines, including C II, as proposed by \citet{san13}.
Finally, a strong emission feature is detected at $\sim$ 7300 \AA, that is identified as [Ca II] $\lambda\lambda$7291-7324 \citep[possibly blended with C II $\lambda$7234 
and He I $\lambda$7281, see][]{san13}, growing in intensity.

\subsection{Spectral sequence of SN 2011hw}  

SN 2011hw was extensively  monitored in optical spectroscopy. The follow-up campaign started  soon after the SN discovery, and  lasted about 70 days. The collection of
spectra is displayed in Figure \ref{sn11hw_spec}. The spectra show  a modest evolution during the entire observational period,
confirming the late discovery of SN 2011hw.  The strongest features are the He I lines, showing complex profiles with  broader and  narrower components (see below). A double-component H$\alpha$ is also detected,
though quite weak. Other Balmer lines, usually prominent in Type IIn SNe, are weak in SN 2011hw. 
The spectrum is dominated by a pseudo-continuum bluewards of $\sim$5600 \AA, and relatively narrow emission features
more prominent in the red spectral region. The nature of the blue pseudo-continuum was widely discussed by \citet{tur93}, \citet{smi12} and \citet{max12}, who suggested that it is the result of the blending
of a forest of narrow and intermediate-width Fe lines, as in the case of SN 2006jc \citep[$v_{FWHM} \approx$ 2000-2500 km s$^{-1}$,][]{smi08,chu09} or SN 2005ip \citep[$v_{FWHM} \approx$ 150-200 km s$^{-1}$,][]{smi09,max12}. 
These lines might explain at the same time 
the apparent step in the continuum at $\sim$5600 \AA, the broad ``W''-shape feature at 4600-5200 \AA~(but also some He I lines may contribute),
and the broad bump between 6100 \AA~and 6600 \AA. \citet{smi12}  noted a major property that distinguishes SN 2011hw spectra from those of SN 2006jc: the presence of 
narrow, high-ionization circumstellar lines.
A comprehensive line identification is shown in Figure \ref{lineID11hw}, where a spectrum of SN 2011hw (December 17, 2011) is shown along with those of two interacting, H-rich SNe, viz. the Type IIn SNe 1995N \citep{pasto05}
and 1988Z \citep{tur93}. We used the line identification performed by \citet{fra02} and \citet{tur93} as guides for the identification of the metal lines in our SN 2011hw spectrum.
We confirm the detection of many high-ionization  lines of \citet{smi12}, including [Ne IV] $\lambda$4714, $\lambda$4716, $\lambda$4726, [Ar X] $\lambda$5536,  
[Ar V] $\lambda$6435 and 
a number of [Fe IV], [Fe V] [Fe VI] and [Fe VII] lines. Other coronal lines clearly detected in SN 2005ip are  weak or absent in SN 2011hw \citep{smi12}. 
A weak, narrow [N II] $\lambda$5765 feature is also visible. The line doublet of [O III] $\lambda\lambda$4959,5007, occasionally
 identified in  spectra of interacting SNe, is possibly seen also in SN 2011hw,  whilst the detection of [O I] and [O II] lines (which are common in Type IIn SNe, though with different strengths, being 
prominent in SN 1995N and weaker in SN 1988Z) is not unequivocal.

\begin{figure*}
\begin{center}
   \includegraphics[angle=270,width=7.0in]{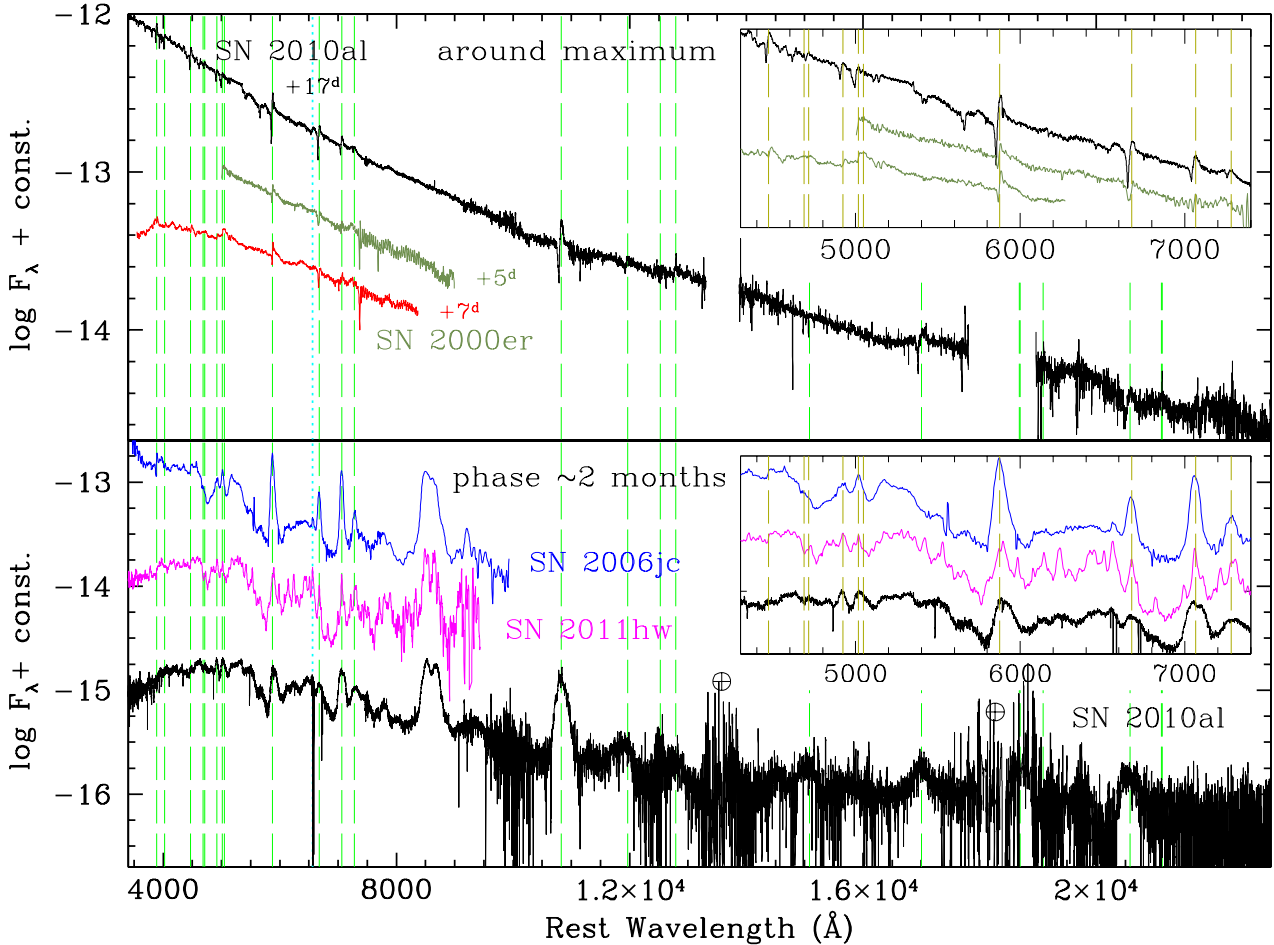} 
\caption{Comparison of early-time ({\bf top}) and $\sim$ 2-months-old ({\bf bottom}) spectra of SN 2010al with those of other Type Ibn SNe at similar phases.
The phases labelled in the upper panel are from the epoch of the core-collapse.
The rest frame positions of the strongest He I lines are marked with green dashed vertical lines. The cyan short-dashed line marks the position of H$\alpha$.
 The inserts show the regions where the most prominent He I features visible in the spectra are located, viz. between 4300 \AA~ and 7400 \AA.}
\label{cfr_specIbn}
\end{center}
\end{figure*}

\citet{smi12} attributed the presence of coronal lines to high-energy photons produced
in shocked material that were able to penetrate the CSM up to the external unshocked regions.
The higher resolution spectra presented by \citet[][see their Figure 5]{smi12} provide evidence for the presence of narrow He I components on top of broader wings. In our highest resolution spectrum 
of November 29 obtained with CAFOS, the narrow lines are visible but not fully resolved. We measure a FWHM velocity of $<$220 km s$^{-1}$ for both He I lines and H$\alpha$, while the intermediate components
extend to about 2000 km s$^{-1}$. Our measurements are in excellent agreement with those of \citet{smi12}. Intermediate components of Mg I (e.g. $\lambda$4571, $\lambda\lambda$5167-5184, $\lambda$7659, 
$\lambda$8213, $\lambda$8924, $\lambda$9256 and $\lambda\lambda$9415-9438), Mg II ($\lambda$4481, $\lambda\lambda$7877-7896 and $\lambda\lambda$9218-9244), O I ($\lambda$7774, $\lambda$8222, 
$\lambda$8446, $\lambda\lambda$9261-9266) and Ca II in the NIR region are clearly detected. Some of these lines 
 possibly show narrow components.

The intermediate components probably arise from shocked material, while the narrow components are linked to the unshocked CSM \citep{smi12}. Consequently, the width of the narrow 
components gives some clues on the velocity of the pre-SN stellar wind. In SN 2011hw, the narrowest spectral lines have $v_{FWHM}$ $\sim$ 200 km s$^{-1}$, which is about one order of magnitude 
lower than the lowest velocity component measured in SN 2006jc. The former value is
unusually low for a typical Wolf-Rayet wind, such as that observed in the CSM of SN 2006jc \citep{pasto07,fol07,smi08}. This, in combination with the clear detection of H in the spectra, 
supports one of the main conclusions of  \citet{smi12}, that the progenitor of SN 2011hw was not a proper Wolf-Rayet star, but probably a transitional object that retained some LBV properties.

\subsection{Comparison of Type Ibn SN spectra}  

In Figure \ref{cfr_specIbn} (top)\footnote{A similar figure was shown in \citet{pasto12}, but without the SN 2011hw spectrum shown here in the bottom panel.} an early-time spectrum of SN 2010al is compared with 
early optical spectra of the Type Ibn  SN 2000er \citep{pasto08a}. The early spectroscopic similarity of the two objects is evident.
The most important He I lines are marked with vertical dashed green lines, the strongest being $\lambda$4471, $\lambda$5876, $\lambda$6678, $\lambda$7065, $\lambda$10830 (the most prominent He I feature), 
$\lambda$17002, $\lambda\lambda$18685-18697 (though contaminated by a wide telluric absorption band) and  $\lambda$20581. H$\alpha$, with a P-Cygni profile, is barely detected in these early spectra.
In Figure \ref{cfr_specIbn} (bottom) a late spectrum of SN 2010al is compared with spectra of SN 2011hw and 2006jc \citep{pasto07}
at similar phases. The spectra of the three objects share a very similar blue pseudo-continuum and the most prominent broader spectral lines. 
However, there are some subtle differences. In particular, H$\alpha$ is still visible as a pure emission -though quite weak- in SN 2011hw,
while it is not unequivocally detected in the late-time spectra of SNe 2006jc and 2010al.

Although there is some heterogeneity in Type Ibn SNe, in general these objects show similar spectral properties and evolution, both in terms of the overall shape of the pseudo-continuum and in
the line identification, profiles and velocities, suggesting that the physical conditions in the line-forming gas regions are not significantly different among the SNe of our sample. All of this 
would indicate a rather similar composition and final configuration in the progenitors of Type Ibn SNe.

\section{Discussion and summary}  \label{disc}

\begin{figure}
\begin{center}
   \includegraphics[angle=0,width=3.4in]{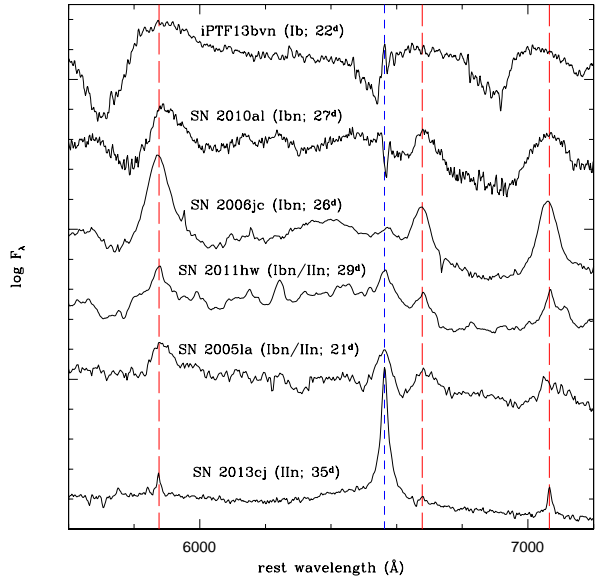} 
\caption{Comparison of a sample of SN spectra obtained a few weeks after their maximum. The phases from maximum are reported in brackets.
The sample includes: the Type Ib SN iPTF13btv \citep{sri14}, the Type Ibn SNe 2010al and 2006jc, the Type Ibn/IIn SNe 2011hw and 2005la and the Type IIn SN 2013cj 
(from the Padova-Asiago SN Archive). Dashed red lines mark the position of strong He I features, the short-dashed blue line marks the position of H$\alpha$.}
\label{cfr_specIbtoIIn}
\end{center}
\end{figure}

Type Ibn SNe represent a small sub-group of CC SNe whose ejecta appear to interact with a dense He-rich, H-poor circumstellar shell. From this point of view, the relative H-to-He abundance in
the CSM and the different velocities inferred from the narrow lines (usually $<$10$^3$ km s$^{-1}$ in SNe IIn, 2-4$\times$10$^3$ km s$^{-1}$ in SNe Ibn) help to discriminate the observed properties of SNe Ibn from those 
of classical SNe IIn \citep[see e.g.][]{tad13}. However, SN 2011hw and, even more, SN 2005la \citep{pasto08b} are quite peculiar in our sample. They show  a non-monotonic post-maximum light curve decline and clear signatures of narrow Balmer
lines in their spectra, suggesting that they are transitional objects between the two CC SN sub-types. Another object, iPTF13beo, has a double peak in the light curve \citep{gor14}, but it does not show H
lines in the spectra.\footnote{A narrow H$\alpha$ has been detected in the spectra of iPTF13beo, but attibuted to galaxy background \citep{gor14}.}
The observational evidence links SNe Ibn to Wolf-Rayet stars, and SNe IIn
(or, at least, some of them) to LBVs \citep[e.g.][]{kot06,smi06,gal07,tru08,tru09,kie12,gro13a,gro13b}. In this context, both SN 2011hw and SN 2005la can be considered as the result of the explosion of 
massive stars that were transiting from the LBV to the WR stages \citep{pasto08b,smi12}.\footnote{We note that \protect\citet{smi12} cast some doubts on the membership of SN 2005la in the Type Ibn SN family, 
because of the slightly different shape of the continuum and the different line ratios. This was because SN 2005la still showed quite strong H lines in the spectrum, together with He lines.  
However, the H/He ratio shown by SN 2005la is very unusual for a Type IIn SN. In addition, the FWHM velocities of the He I lines
\citep[$\sim$ a few $\times$ 10$^3$ km s$^{-1}$,][]{pasto08b} are closer to those expected in WR winds rather then those one can find in typical LBV winds
\citep{abb87,hum94}.} \citet{smi12} provided robust evidences of the link of
SN 2011hw to early WN stars with some residual H, or even Opfe/WN9 stars (although the wind velocity observed in the latter is significantly lower than that observed in the CSM of SN 2011hw).
In this case, if the progenitor star belongs to a massive binary system, the mass transfer from the companion might enable the LBV to WN or Opfe/WN9 transition.
As an intriguing consequence, SNe Ibn might be regarded as transitional objects in a sort of sequence 
between WR-related events (some Type Ib/c SNe) and LBV-related events (some SNe IIn). The existence of a continuity from SNe IIn to stripped-envelope SNe is
illustrated in the comparison among post-maximum spectra in Figure \ref{cfr_specIbtoIIn}. Although we admit that this conclusion is speculative, this possibility
should be explored by carefully analysing a wider Type Ibn SN sample. A more detailed 
discussion on the variety of properties observed in Type Ibn SNe, including  the study of the evolution of a range of observed parameters (e.g
the velocity of the different line components), will be faced in a forthcoming paper (Pastorello et al. in preparation).

SN 2006jc and similar objects (e.g. SNe 1999cq and 2002ao), which we indicate as prototypical SNe Ibn, 
show very weak or even no evidence of H in the CSM \citep{mat00,pasto07,fol07,pasto08a}. 
This points toward progenitors that have lost most of their He-rich external layers, and were stripped of their H envelopes a long time before their CC SN explosion. 
This would therefore connect SN 2006jc to WN or even WCO progenitors \citep{tom08}. In this context, SN 2010al shows some differences. The early-time spectrum 
was dominated by a blue continuum, relatively prominent Balmer features, He II lines with $v_{FWHM} \approx$ 2250 km s$^{-1}$, 
and there was evidence for the presence of lines of CNO processing material (in particular the blend of N III $\lambda$4640 and C III $\lambda$4648, see Section \ref{sec:spec10al}).
Such lines have been seen in the CSM of the Type IIn SNe 1998S \citep{fas01,fra05} and 2008fq \citep{tad13}, and considered indicative of CNO-element enrichment in the circumstellar wind
produced by the progenitor star. This suggested enhanced mixing from the inner CNO burning regions to the outer layers, possibly favoured by stellar rotation.
In the case of SN 2010al, the identified circumstellar lines and their velocities (exceeding 10$^3$ km s$^{-1}$) are well consistent with what expected in winds
of WR stars.

A few days later, the spectrum of SN 2010al experienced a significant evolution, showing marginal or no evidence of He II and CNO element lines. It was instead characterized by
 narrow lines of He I (with an expansion velocity of about 1000 km s$^{-1}$ from the position of the P-Cygni absorption minimum), whilst there was only a marginal evidence for the presence of H$\alpha$.
At later epochs the spectra displayed the characteristic blue pseudo-continuum that is observed in SNe Ibn (and other interacting SNe), with the He I features showing a broader P-Cygni profile. 
The FWHM velocity of the He I increased to about 5-6 $\times$ 10$^3$ km s$^{-1}$, and became comparable with that measured for the strong Ca II and 
O I lines. Since these $\alpha$-element lines are believed to form in the SN ejecta, this is probably an indication that a significant amount of He was still present in the pre-SN stellar envelope.
In other words, the pre-SN star was probably still He-rich at the time of the collapse of the core. This links SN 2010al to more canonical
type Ib SNe. The similarity with SNe Ib can also be noticed in the early photometric evolution of SN 2010al: the overall light curve shape is reminiscent of those of normal 
SNe Ib\footnote{SN 1999cq, here taken as representative for typical SNe Ibn, showed a rather asymmetric light curve with a very sharp rise to maximum and a relatively fast decline in the optical bands. OGLE-2012-SN006 \citep{pasto13b}
had a photometric evolution similar to that of SN 2010al in the pre-maximum phase and during the first month past maximum. However, later on, the light curve of OGLE-2012-SN006 significantly flattened. This behaviour has never been observed in Type Ibn SNe before.},
although with a much wider post-peak magnitude drop ($\Delta M$ $\sim$ 5 mag in about 40 days). 
 This would suggest that the luminosity from ejecta-CSM interaction contributes mostly to the very luminous light curve peak.

A plausible explanation for the evolution of the He I  lines is that it is determined by the properties of the CSM:  evidence of the underlying SN ejecta can be inferred from the apparent broadening of 
the He I lines that would be indicative of the growing emission contribution from the shocked ejecta. In this context, it is relevant here to discuss the outcomes of the X-ray observations (0.2-10 keV)
of a sample of interacting SNe 
presented by \citet{ofe13}, since three Type Ibn (including the two objects presented in this paper) were included in the SN sample of Ofek et al.
They found that two Type Ibn SNe (SN 2006jc and 2010al) had very similar peak X-ray luminosities (L$_X \approx$2 $\times$ 10$^{41}$ erg s$^{-1}$), whilst for SN 2010hw they inferred an 
upper luminosity limit L$_X <$ 5 $\times$ 10$^{40}$ erg s$^{-1}$.  In most cases of interacting SNe, the X-ray emission arises from the diffusion of shock breakout energy in a circumstellar shell.
However, this explanation does not match the case of SN 2010al. Usually, in fact, the intensity of the X-ray emission from  shock breakout energy diffusion is lower than that originating
from ejecta-CSM interaction. In addition, \citet{ofe13} determined a surprisingly high L$_X$/L$_{opt}$ ratio for SN 2010al, about 0.3, which 
is one order of magnitude higher than that measured for SN 2006jc (and about 2 orders of magnitude more than that estimated for SN 2011hw). Finally, the maximum of the L$_X$ light curve
is reached quite early (about 15 days from the optical maximum estimated to occur on JD = 2455284.3).\footnote{We note that the epoch of the optical maximum for SN 2010al computed in this paper
is over 2 weeks later than that assumed in the \protect\citet{ofe13} paper (JD = 2455268.5).} All of this makes implausible the shock breakout-CSM scenario for SN 2010al.
More plausible is that the high X-ray emission of SN 2010al  is determined by the conversion of a large fraction of kinetic energy into radiation, via shock-interaction of the SN ejecta with a 
circumstellar shell expelled a short time before the SN explosion. We remark that in coincidence with the peak of X-ray emissivity, the optical spectrum changes significantly, and becomes dominated by
intermediate-width features.

In the case of SN 2011hw, the spectra show a variety of lines with multi-component profiles, and clear evidence of narrow, high-ionization forbidden Fe lines \citep{smi12}. These are 
signatures that high-frequency
radiation ionizes the outer unshocked CSM. Taking into account the peculiar, non-linear decline of the optical light curve of SN 2011hw, the implication is that the SN ejecta interact with a CSM 
having a rather complex density profile and this likely powers the evolution of SN 2011hw for a longer period than for SN 2010al. We note that structured density profiles are quite common in the
CSM of interacting SNe \citep[see, e.g.,][]{zam05}.

The spectral and photometric properties of SN 2010al indicate that the best scenario for this object is that of a Type Ib SN exploded in a He-rich, H-poor CSM. This implies that the progenitor star was totally
H-stripped, and exploded while it was still losing the residual He-rich layers. The best candidate precursor of this SN is a WR star (probably a WN Type). 
The observables of SN 2011hw, on the other hand,  collocate this  object as a transitional event between a Type IIn SN and a Type Ibn SN. We agree with the conclusions of \citet{smi12} that the precursor of SN 2011hw
was a post-LBV star, possibly an Ofpe/WN9 star, with some residual H in the envelope. In this sense the two objects may represent the extremes of the Type Ibn SN variety, and  -with
other Type Ibn events \citep[see Table 4 in][for a comprehensive list]{pasto13}- they contribute to fill the {\it observational} gap between Type IIn and stripped-envelope SNe.

The Type Ibn SNe analyzed so far confirm that they are produced by WR stars, and the collapse of the stellar core may occasionally be heralded by major mass-loss events occurring 
a short time before the terminal SN explosion, as directly observed in SN 2006jc.
The heterogeneity in the observed parameters of SNe Ibn are likely related to the different physical, geometrical and chemical properties of the CSM. 
This possibly results from different evolutionary paths for Type Ibn SN progenitors, implying that stars with a range of initial masses may reach 
different stages in the post-LBV stellar evolution toward a stripped C-O core.

\section*{Acknowledgments}

AP thanks Elia Cozzi, Rafael Benavides $\&$ Joseph Brimacombe for kindly providing their observations of SN 2010al for this study, and M. L. Pumo for useful discussions.
We would like to thank the anonymous referee for providing us with constructive comments and suggestions.

AP, EC, SB, MLP, AH, LT, and MT are partially supported by the PRIN-INAF 2011 with the project "Transient Universe: from ESO Large
 to PESSTO". DYT is partly supported by the Russian Foundation for Basic Research (project No. 13-02-92119).
 FB acknowledges support from FONDECYT through Postdoctoral grant 3120227 and by the Millennium Center for Supernova Science through                                       
grant P10-064-F (funded by "Programa Bicentenario de Ciencia y Tecnolog\'ia                                                       
de CONICYT" and "Programa Iniciativa Cient\'ifica Milenio de MIDEPLAN").  
ST acknowledges support by the Transregional Collaborative Research Centre TRR 33 of the German Research Foundation.                                                        
MDS gratefully acknowledges generous support provided by the Danish Agency for Science and Technology and Innovation  
realized through a Sapere Aude Level 2 grant. 
XW is supported by the Major State Basic Research Development Program (2013CB834903), 
the National Natural Science Foundation of China (NSFC grants 11073013, 11178003, 11325313), and the Foundation of Tsinghua University (2011Z02170)
This work is supported in part under grant number 1108890 from the US National Science Foundation and is partly supported by the European Union FP7 
programme through ERC grant number 320360. 
This material is based upon work supported by NSF under grants AST--0306969, AST--0607438 and AST--1008343.                            

The MASTER network has made use of equipment acquired through the Development Program for Lomonosov Moscow State University, and was
also supported by a grant in the form of a subsidy from the Ministry of Education and Science of the Russian Federation
(agreement of August 27, 2012, No. 8415) and by the "Dinastiya" Foundation for Non-Commercial Programs.
The work is supported by the Russian Foundation of Fundamental Research, grant RFFI 14-02-31546.

This work is partially based on observations of the European supernova collaboration involved in the ESO-NTT large programme 184.D-1140 led by Stefano Benetti. 
It is also based on observations made with ESO VLT Telescopes at the Paranal Observatory under program IDs 084.D-0265 and 085.D-0701 (PI. S. Benetti).
The manuscript includes data obtained at the Large Binocular Telescope of the Mount Graham International Observatory.
The LBT is an international collaboration among institutions in the United States, Italy and Germany. LBT Corporation partners are: The University of Arizona 
on behalf of the Arizona university system; Istituto Nazionale di Astrofisica, Italy; LBT Beteiligungsgesellschaft, Germany, representing the Max-Planck Society, 
the Astrophysical Institute Potsdam, and Heidelberg University; The Ohio State University, and The Research Corporation, on behalf of The University of Notre Dame, 
University of Minnesota and University of Virginia.

  This paper is based on observations made with the Italian Telescopio Nazionale Galileo 
(TNG) operated on the island of La Palma by the Fundaci\'on Galileo Galilei of         
the INAF (Istituto Nazionale di Astrofisica). It is also based on observations made with 
the Liverpool Telescope (LT) operated on the island of La Palma at the Spanish Observatorio del                                             
Roque de los Muchachos of the Instituto de Astrofisica de Canarias; the William Herschel Telescope (WHT) 
operated on the island of La Palma by the Isaac Newton Group in the Spanish Observatorio del Roque de los Muchachos of the Instituto de Astrofísica de Canarias; 
the Nordic Optical Telescope (NOT), operated on the island of La Palma jointly by Denmark, Finland, Iceland,                                                               
Norway, and Sweden, in the Spanish Observatorio del Roque de los Muchachos of the Instituto de Astrof\'isica de Canarias; 
the 0.80m  Tsinghua-NAOC Telescope (TNT) in the Xinglong Observatory of NAOC;
the 2.2m Telescope of  the Centro Astron\'omico Hispano Alem\'an    
(CAHA) at Calar Alto, operated jointly by the Max-Planck Institut f\"ur                                                       
Astronomie and the Instituto de Astrof\'isica de Andaluc\'ia (CSIC); the 1.82m Copernico Telescope and the 67/92-cm Schmidt Telescope of                         
the INAF-Asiago Observatory; the 2-m Faulkes Telescope North at Haleakala Observatory
(Hawaii, US). This work makes use of observations from the LCOGT network.

This publication makes use of data products from the Two Micron All Sky Survey, which is a joint project of the University of Massachusetts and the Infrared Processing and Analysis Center/California Institute of Technology, 
funded by the National Aeronautics and Space Administration and the National Science Foundation.

\appendix

\section[]{Photometry Tables}

\begin{table*}
\caption{Optical magnitudes of the sequences of local standards in the fields of SNe 2010al and 2011hw (Figure \ref{fig:app_maps}). The errors are the rms of the recovered magnitudes. If no error is indicated in brackets,
the reported magnitude is that of a single (plausibly photometric) night.}
\begin{center}
%\small
\begin{tabular}{cccccc} \hline\hline
\multicolumn{6}{c}{SN 2010al} \\ \hline
Star & $U$ & $B$ & $V$ & $R$ & $I$ \\ \hline
s1  &     --         &  18.09 (0.03) & 17.45 (0.04)  & 17.01 (0.07)  & 16.43 (0.07)  \\
s2  &    13.69       &  13.72 (0.03) & 13.17 (0.03)  & 12.85 (0.07)  & 12.49 (0.07)  \\
s3  &     --         &     --        & 18.44 (0.14)  & 17.85 (0.07)  &     --        \\
s4  &     --         &  19.77 (0.11) & 18.48 (0.09)  & 17.58 (0.02)  & 16.77 (0.02)  \\
s5  & 13.40 (0.07)   &  13.48 (0.03) & 13.02 (0.02)  & 12.80 (0.06)  & 12.47 (0.02)  \\
s6  &       --       &     --        & 19.79 (0.07)  & 18.47 (0.02)  & 17.04 (0.02)  \\
s7  &   16.12        &  15.86 (0.03) & 15.07 (0.02)  & 14.66 (0.04)  & 14.24 (0.02)  \\
s8  & 16.82 (0.05)   &  16.80 (0.03) & 16.27 (0.02)  & 15.89 (0.02)  & 15.55 (0.05)  \\
s9  & 16.80 (0.07)   &  16.15 (0.02) & 15.25 (0.02)  & 14.76 (0.07)  & 14.21 (0.04)  \\
s10 & 15.78 (0.03)   &  15.85 (0.03) & 15.30 (0.03)  & 14.99 (0.04)  & 14.60 (0.04)  \\
s11 &   15.50        &  15.50 (0.03) & 14.86 (0.03)  & 14.48 (0.06)  & 14.09 (0.06)  \\
s12 &   14.94        &  15.00 (0.03) & 14.46 (0.03)  & 14.12 (0.07)  & 13.89 (0.09)  \\
s13 &   15.96        &  15.40 (0.03) & 14.54 (0.03)  & 14.02 (0.04)  & 13.57 (0.06)  \\
s14 &   20.13        &  18.98 (0.03) & 17.46 (0.02)  & 16.69 (0.06)  & 15.76 (0.05)  \\
s15 &   17.32        &  17.35 (0.04) & 16.84 (0.04)  & 16.50 (0.07)  & 16.07 (0.06)  \\
s16 &     --         &     --        & 18.99 (0.07)  & 17.88 (0.07)  & 16.59 (0.07)  \\
\hline\hline
\multicolumn{6}{c}{SN 2011hw} \\ \hline
Star &  $U$ & $B$ & $V$ & $R$ & $I$   \\ \hline
s1 &  12.82 (0.01) & 12.69 (0.01) & 12.08 (0.01) & 11.73 (0.01) & 11.39 (0.01) \\
s2 &  14.80 (0.01) & 14.83 (0.01) & 14.29 (0.01) & 13.94 (0.01) & 13.58 (0.01) \\
s3 &  15.32 (0.02) & 15.22 (0.01) & 14.51 (0.01) & 14.09 (0.01) & 13.68 (0.01) \\
s4 &  16.67 (0.02) & 16.50 (0.01) & 15.80 (0.02) & 15.39 (0.01) & 14.99 (0.01) \\
s5 &   --          & 16.04 (0.02) & 15.36 (0.04) & 14.85 (0.05) & 14.50 (0.04) \\
s6 &  15.43 (0.01) & 15.42 (0.01) & 14.81 (0.01) & 14.43 (0.01) & 14.07 (0.01) \\
s7 &  16.64 (0.01) & 16.44 (0.01) & 15.73 (0.01) & 15.29 (0.01) & 14.89 (0.01) \\
s8 &  16.84 (0.03) & 16.73 (0.01) & 16.08 (0.02) & 15.70 (0.02) & 15.33 (0.01) \\
s9 &  17.69 (0.03) & 17.19 (0.02) & 16.32 (0.01) & 15.83 (0.01) & 15.36 (0.01) \\
\hline
\end{tabular}
\end{center}
\label{tab_seqstars}
\end{table*}

\begin{table*}
\caption{Swift/UVOT $UV$-bands magnitudes of the reference stars in the fields of SNe 2010al and 2011hw. 
The errors are the rms of the recovered magnitudes.}
\begin{center}
%\small
\begin{tabular}{ccccccc} \hline\hline
\multicolumn{4}{c}{SN 2010al} \\ \hline
Star & $uvw2$ & $uvm2$ & $uvw1$ & $u$ & $b$ & $v$ \\ \hline
s5  & 15.24 (0.03) & 14.94 (0.03) & 14.23 (0.03) & -- & -- & -- \\
s7  & 19.20 (0.14) & 19.98 (0.21) & 17.55 (0.08) & 16.01 (0.05) &15.73 (0.04) &15.02 (0.04) \\
s8  & 19.44 (0.15) & 19.38 (0.15) & 18.03 (0.10) & 16.84 (0.07) &16.79 (0.05) &16.27 (0.07) \\
s9  & 19.70 (0.19) &    --        & 18.34 (0.12) & 16.67 (0.07) &16.17 (0.04) &15.30 (0.04) \\
s10 & 18.15 (0.08) & 17.94 (0.07) & 16.88 (0.06) & 15.70 (0.04) &15.78 (0.03) &15.36 (0.05) \\
s11 & 18.08 (0.09) & 17.95 (0.08) & 16.63 (0.06) & 15.37 (0.04) &15.39 (0.03) &14.90 (0.04) \\
s12 & 16.99 (0.06) & 16.66 (0.05) & 15.88 (0.05) & 14.83 (0.04) &14.94 (0.03) &14.51 (0.04) \\
s13 & 19.08 (0.13) & 20.02 (0.22) & 17.46 (0.08) & 15.78 (0.05) &15.33 (0.03) &14.56 (0.04) \\
s15 & 19.76 (0.19) & 19.45 (0.16) & 18.49 (0.13) & 17.27 (0.09) &17.32 (0.07) &16.81 (0.09) \\
\hline\hline
\multicolumn{4}{c}{SN 2011hw} \\ \hline
Star & $uvw2$ & $uvm2$ & $uvw1$ & $u$ &  $b$ & $v$ \\ \hline
s1 & 15.73 (0.03) & 15.88 (0.03) &  14.16 (0.03) & 12.77 (0.03) & 12.74 (0.03) & 12.12 (0.02) \\  
s2 & 17.23 (0.04) & 16.93 (0.04) &  15.87 (0.03) & 14.73 (0.04) & 14.81 (0.02) & 14.30 (0.03) \\  
s3 & 18.31 (0.06) & 18.61 (0.07) &  16.70 (0.04) & 15.26 (0.04) & 15.17 (0.03) & 14.52 (0.04) \\
s4 & 19.64 (0.12) & 20.01 (0.16) &  17.99 (0.07) & 16.59 (0.04) & 16.48 (0.02) & 15.83 (0.04) \\
s5 & 19.38 (0.11) & 20.14 (0.17) &  17.82 (0.06) & 16.25 (0.05) & 16.08 (0.03) & 15.35 (0.03) \\
s6 & 18.27 (0.06) & 18.23 (0.06) &  16.74 (0.04) & 15.39 (0.04) & 15.41 (0.02) & 14.84 (0.03) \\
s7 & 19.69 (0.13) & 20.04 (0.17) &  18.13 (0.07) & 16.62 (0.05) & 16.47 (0.03) & 14.79 (0.03) \\
s8 & 19.69 (0.13) & 19.60 (0.12) &  18.12 (0.07) & 16.78 (0.05) & 16.71 (0.03) & 16.11 (0.04) \\    
\hline
\end{tabular}
\end{center}
\label{tab_seqstars_UV}
\end{table*}

\begin{table*}
\caption{Calibrated  photometry of SN 2010al. The errors in brackets are obtained by combining in quadrature the errors in the photometric calibration and 
instrumental PSF measurement errors. The symbol ``$^\ast$'' marks unfiltered data rescaled to the $R$-band magnitudes; the symbol ``$^\dag$'' indicates 
luminance filter measurements reported to $V$-band magnitudes. Standard Johnson-Cousins $V$- and $R$-band magnitudes from amateur astronomers have been obtained by 
computing instrumental zeropoints using the $V$ and $R$ magnitudes of the local sequence stars reported in Table \ref{tab_seqstars}, 
and adopting no colour correction. The numbers in the last column identify the intrumental configurations (see table footnotes).
A relative uncertainty of 10 per cent on the flux calibration has been assumed for the VLT spectra.
}
\begin{center}
\tiny
\begin{tabular}{cccccccccccc} \hline\hline
Date & $JD$ & $U$ & $B$ & $V$ & $R$ & $I$ & $J$ & $H$ & $K$ & Source \\ 
   & $+$2455000 & & & & & & & & \\ \hline
07Feb10$^\ast$ & 234.64    &  -- &   -- & -- &   $>$19.83      &  -- & -- & -- & -- &  1 \\
12Mar10       & 268.21    &  -- &   -- & -- &   $>$19.16      &  -- & -- & -- & -- &  2 \\ %Kislov 
13Mar10$^\ast$ & 268.53    &  -- &   -- & -- &    18.87  (0.24) & -- & -- & -- & -- &  1 \\
15Mar10$^\ast$ & 271.39    &  -- &   -- & -- &    17.65  (0.27) & -- & -- & -- & -- &  3 \\
16Mar10$^\ast$ & 271.57    &  -- &   -- & -- &    17.57  (0.25) & -- & -- & -- & -- &  1 \\
16Mar10$^\dag$ & 271.93    &  -- &   -- &    17.34  (0.13) & -- & -- & -- & -- & -- &  4 \\
16Mar10       & 271.95    &  -- &   -- & -- & 17.35  (0.32)    & -- & -- & -- & -- &  5 \\        
16Mar10$^\dag$ & 272.01    &  -- &   -- &    17.33  (0.10) & -- & -- & -- & -- & -- &  4 \\
16Mar10       & 272.43    &  -- &   -- & -- &   17.00  (0.08)  & -- & -- & -- & -- & 2 \\ %Kislov      
17Mar10$^\ast$ & 272.54    &  -- &   -- & -- &   16.94  (0.19)  & -- & -- & -- & -- &  1 \\
17Mar10       & 272.97    &  -- & 16.77 (0.09) & 16.97 (0.06) & 16.77 (0.08) & 16.70 (0.08) &  -- & -- & -- & 6 \\ 
17Mar10       & 273.01    &  -- &   -- & -- &   16.75  (0.12)  & -- & -- & -- & -- & 5 \\ %Blagov      
17Mar10       & 273.39    &  -- &   -- & -- &   16.71  (0.10)  & -- & -- & -- & -- & 2 \\ %Kislov      
18Mar10$^\ast$ & 273.50    &  -- &   -- & -- &   16.66  (0.09)  & -- & -- & -- & -- &  1  \\
%18Mar10      & 274.04    & -- &   -- & -- &   16.22  (0.11)  & -- & -- & -- & -- &  5 \\ Blagov      
18Mar10       & 274.19    &  -- &   -- & -- &   16.46  (0.05)  & -- & -- & -- & -- &  2 \\ %Kislov      
19Mar10       & 275.06    &  -- &   -- & -- &   16.23  (0.10)  & -- & -- & -- & -- & 5  \\ %Blagov      
19Mar10       & 275.21    &  -- &   -- & -- &   16.25  (0.05)  & -- & -- & -- & -- & 2 \\ %Kislov      
20Mar10       & 276.36    &  -- &   -- & -- &   15.98  (0.06)  & -- & -- & -- & -- & 2 \\ %Kislov      
21Mar10       & 277.23    &  -- &   -- & -- &   16.00  (0.06)  & -- & -- & -- & -- & 2 \\  %Kislov      
22Mar10       & 278.25    &  -- &   -- & -- &   15.74  (0.02)  & -- & -- & -- & -- & 2 \\  %Kislov    %%%%%%%%%%%%  
23Mar10       & 279.49    &  -- &   -- & -- &  -- & -- & 15.35 (0.42) &  15.32 (0.50) & -- & 7\\
25Mar10       & 280.55    & 15.08  (0.12) & 15.91 (0.12) & 15.83 (0.12) & 15.72 (0.12) & 15.66 (0.12) & 15.33 (0.12) & 15.19 (0.12) & 15.05 (0.12) & 8\\  %VLT
26Mar10       & 282.04    &  -- &  -- & 15.84 (0.06) & 15.69 (0.09) & -- &  -- & -- & -- & 6 \\ 
26Mar10       & 282.11    &  -- & 15.93 (0.05) & 15.84 (0.06) & 15.71 (0.04) & 15.63 (0.08) &  -- & -- & -- & 6 \\ 
27Mar10       & 282.50    &  -- &   -- & -- &  -- & -- & 15.31 (0.45) &  15.18 (0.31) & -- & 7\\
27Mar10$^\ast$ & 282.60    &  -- &   -- & -- &   15.63  (0.05)  & -- & -- & -- & -- & 1 \\
27Mar10       & 283.08    &  -- & 15.95 (0.15) & 15.85 (0.06) & 15.70 (0.05) & 15.57 (0.06) &  -- & -- & -- & 6 \\ 
28Mar10       & 283.97    &  -- & 15.95 (0.14) & 15.82 (0.07) & 15.66 (0.08) & 15.53 (0.12) &  -- & -- & -- & 6 \\ 
28Mar10       & 283.99    &  -- &   -- & -- &   15.56  (0.11)  & -- & -- & -- & -- &  5\\  %Blagov      
29Mar10       & 284.50    & 15.21 (0.12) &  15.93 (0.12) & 15.80 (0.12) & 15.70 (0.12) & 15.57 (0.12) & 15.46 (0.12) &  15.33 (0.12) & 15.37 (0.12) & 8 \\ %VLT
02Apr10$^\ast$ & 289.44    &   -- &   -- & -- &   15.65 (0.07)  & -- & -- & -- & -- &  9 \\
04Apr10       & 291.37    &  -- &   -- & -- &  -- & -- & 15.77 (0.38) &  15.34 (0.32) & -- & 7\\
05Apr10$^\ast$ & 291.57    &   -- &   -- & -- &   15.77 (0.07)  & -- & -- & -- & -- &  1 \\
06Apr10       & 292.97    &   -- &   -- & -- &   15.94 (0.08)  & -- & -- & -- & -- &  5 \\ %Blagov      
06Apr10       & 292.99    &  -- & 16.44 (0.05) & 16.13 (0.08) & 15.92 (0.05) & 15.68 (0.07) &  -- & -- & -- & 6 \\ 
06Apr10       & 293.45    &  --  &   16.47 (.11) & 16.13 (0.06) & 15.92 (0.09) & 15.72 (0.08)  & -- & -- & -- & 10 \\ %Ekar_tempsub 
07Apr10       & 294.49    & 15.95  (0.12) &  16.47 (0.12) &  16.18 (0.12) & 16.03 (0.12) & 15.80 (0.12) & 15.76 (0.12) & 15.49 (0.12) & 15.16 (0.12) & 8 \\ %VLT
11Apr10$^\ast$& 297.57    &  -- &   -- & -- &    16.31 (0.06)  & -- & -- & -- & -- &  1 \\
11Apr10       & 298.36    &  -- &   -- & -- &  -- & -- & 16.03 (0.24) &  15.79 (0.22) & -- & 7\\
14Apr10       & 300.99    &  -- & 17.35 (0.08) & 16.83 (0.08) & 16.56 (0.06) & 16.20 (0.07) &  -- & -- & -- & 6 \\ 
17Apr10       & 303.55    & 17.85 (0.12) & 17.94 (0.05) & 17.43 (0.05) & 17.03 (0.07) & 16.62 (0.05) & -- & -- & -- &  11 \\ % NTT
19Apr10       & 305.61    &  -- & -- & -- & -- & -- &  16.76 (0.12) & 16.77 (0.18 )& 16.49 (0.14) & 12 \\ %SOFI
22Apr10       & 309.36    &  -- &   -- & -- &  -- & -- & 16.98 (0.40) &  16.96 (0.28) & -- & 7\\
25Apr10       & 312.38    &  -- &   -- & -- &  -- & -- & 17.58 (0.45) &  17.33 (0.46) & -- & 7\\
30Apr10       & 316.67    &  -- & $>$18.58 & 19.36 (0.32) & -- & -- &  -- & -- & -- & 6 \\ 
01May10$^\ast$ & 317.54   &  -- &   -- & -- &    18.76  (0.31) & -- & -- & -- & -- &  1 \\
05May10       & 321.41   &  20.46 (0.36) & 20.24 (0.09) & 19.68 (0.10) & 19.29 (0.15) & 18.76 (0.14) & -- & -- & -- & 13 \\ %TNG
11May10       & 328.49   &  20.94 (0.12) & 20.92 (0.12) & 20.31 (0.12) & 20.16 (0.12) & 19.87 (0.12) & 19.33 (0.12) & 18.65 (0.12) &  18.36 (0.12) & 8 \\ %VLT
14Sep10       & 453.68   &  -- &   $>$22.29  &  $>$22.15  &  $>$21.55  &  $>$21.19 &   -- & -- & -- &  14 \\ %CAHA
11Oct10       & 481.00    &  -- & -- & -- & -- & -- &  20.34 (0.14) & 19.64 (0.12) & 19.60 (0.23) & 15 \\ % LBT
26Oct10       & 495.58    & $>$19.00 & $>$20.59 & $>$19.28 & $>$19.20 & $>$19.68 & -- & -- & -- &  14 \\ %CAHA
02Jan11       & 563.64   & $>$19.98 & $>$20.12 & $>$20.06 & $>$19.87 & $>$19.49 & -- & -- & -- &  11 \\ 
31Jan11       & 592.65   & -- & -- & -- & -- & -- & $>$20.41 & $>$19.74 & -- & 15 \\ % LBT
31Mar11       & 651.63   & -- & -- & -- & -- & -- & $>$19.50 & -- & $>$17.55 & 15 \\ \hline % LBT
\end{tabular}
\\
1 = Meade 16" Reflector + SBIG ST-9XE Dual CCD camera  (Rich Observatory, Hampden, Maine, USA; obs. D.R.); 
2 = 0.4-m MASTER  telescope + Apogee Alta U16M CCD (Kislovodsk, Caucasian region Russia); 
3 = 0.36-m Schmidt-Cassegrain + Apogee ALTA U47 CCD camera (New Millennium Observatory, Mozzate, Italy; obs. E. Cozzi);
4 = 12.5-inch RCOS Telescope + SBIG STL6303 CCD camera (Macedon Ranges Observatory, Melbourne, Australia; obs. J. Brimacombe); 
5 = 0.4-m MASTER  Telescope + Apogee Alta U16M CCD (Blagoveschensk; Far East region, Russia);
6 = 0.80-m Tsinghua-NAOC Telescope + Princeton Instruments VersArray:1300B CCD (Xinglong Observatory, Yanshan mountains, Hebei, China).
7 = 2.0-m Liverpool Telescope + SupIRCam (La Palma, Canary Islands, Spain);
8 = 8.2-m Very Large Ttelescope (UT2) + XShooter (spectro-photometry; European Southern Observatory; Cerro Paranal, Chile);
9 = 280-mm Celestron 11  + Atik 16HR with Sony chip ICX285AL (Posadas Observatory; C\`ordoba, Spain; obs. R. Benavides);
10 = 1.82-m Copernico Telescope + AFOSC (Mt. Ekar, Asiago, Italy); 
11 = 3.58-m New Technology Telescope + EFOSC2 (European Southern Observatory; La Silla, Chile);
12 = 3.58-m New Technology Telescope + SOFI (European Southern Observatory; La Silla, Chile);
13 = 3.58-m Telescopio Nazionale Galileo + Dolores (La Palma, Canary Islands, Spain); 
14 = 2.2-m Calar Alto Telescope + CAFOS (Calar Alto Observatory, Almer\`ia, Spain);
15 = 2 $\times$ 8.4-m Large Binocular Telescope + Lucifer (Mt. Graham International Observatory, Arizona, USA);
%"Rafael Benavides" <rafaelbenpal@gmail.com>
%jbrimaco@bigpond.net.au
%"Doug Rich" <richobservatory@att.net>
\end{center}
\label{tab_ph_10al}
\end{table*}%

\begin{table*}
\caption{Table with the calibrated multi band photometry of SN 2011hw. The errors in brackets are obtained by combining in quadrature the errors of the photometric calibration and the 
instrumental PSF measurement errors. The symbol ``$^\ast$'' indicates unfiltered measurements rescaled to $R$-band magnitudes. These have been obtained by computing zeropoints using the $R$-band magnitudes of the stellar sequence in the SN field, and assuming negligible colour correction.}
\begin{center}
\small
\begin{tabular}{ccccccccc} \hline\hline
Date & $JD+$2455000 & $U$ & $B$ & $V$ & $R$ & $I$  & Source \\  \hline
%   & 2455000 & & & & & & & \\ \hline
19Nov11 &  885.27 &    -- & 16.99 (0.02) & 16.86 (0.03) & 16.60 (0.02) & 16.31 (0.03) &  1 \\
20Nov11 &  886.32 &  16.47 (0.02) & 17.00 (0.03) & 16.85 (0.04) & 16.62 (0.04) & 16.30 (0.05) &  1 \\
21Nov11 &  887.29 &    -- & 17.05 (0.12) & 16.85 (0.19) & 16.68 (0.14) & 16.31 (0.18) & 2 \\
22Nov11 &  888.24 &    -- & 17.14 (0.02) & 16.87 (0.09) & 16.68 (0.05) & 16.37 (0.07) & 2 \\
23Nov11 &  889.30 &    -- & 17.15 (0.04) & 16.87 (0.05) & 16.71 (0.13) & 16.36 (0.14) & 2 \\
24Nov11 &  890.46 &    -- & 17.12 (0.08) & 16.92 (0.19) & 16.81 (0.17) & 16.42 (0.26) &  1 \\
27Nov11 &  893.25 &  16.57 (0.02) & 17.11 (0.06) & 16.99 (0.04) & 16.76 (0.10) & 16.41 (0.04) &  1 \\
29Nov11 &  895.39 &  16.55 (0.03) & 17.11 (0.02) & 16.98 (0.02) & 16.77 (0.02) & 16.32 (0.03) &  3 \\
10Dec11 &  905.72 &  16.19 (0.04) & 16.90 (0.04) & 16.76 (0.05) & 16.58 (0.04) & 16.21 (0.07) &  4 \\
11Dec11 &  907.37 &  16.21 (0.03) & 16.89 (0.02) & 16.79 (0.02) & 16.65 (0.02) & 16.27 (0.02) &  5 \\
17Dec11 &  912.76 &  16.50 (0.04) & 17.02 (0.03) & 16.83 (0.01) & 16.65 (0.05) & 16.34 (0.08) &  4 \\
17Dec11 &  913.39 &  16.60 (0.02) & 17.08 (0.03) & 16.90 (0.02) & 16.71 (0.02) & 16.39 (0.02) &  6 \\ 
18Dec11$^\ast$ &  914.26 &    -- &  -- &  -- & 16.71 (0.16) &  -- & 1 \\
21Dec11 &  917.34 &  16.92 (0.03) & 17.35 (0.01) & 17.06 (0.01) & 16.96 (0.01) & 16.52 (0.01) &  7 \\
22Dec11 &  917.76 &  16.81 (0.06) & 17.34 (0.03) & 17.07 (0.05) & 16.97 (0.07) & 16.56 (0.08) &  4 \\
23Dec11 &  919.26 &  16.98 (0.03) & 17.37 (0.03) & 17.20 (0.02) & 17.08 (0.03) & 16.55 (0.07) &  3 \\
26Dec11 &  922.36 &  17.03 (0.06) & 17.58 (0.02) & 17.30 (0.03) & 17.29 (0.02) & 16.90 (0.02) &  5 \\
29Dec11 &  924.73 &  17.29 (0.05) & 17.70 (0.03) & 17.37 (0.04) & 17.31 (0.02) & 16.98 (0.04) &  4 \\
01Jan12 &  928.27 &    -- & 17.91 (0.03) & 17.64 (0.02) & 17.56 (0.02) & 17.03 (0.04) &  3 \\
04Jan12 &  931.27 &    -- & 18.05 (0.32) & 17.76 (0.30) & 17.69 (0.17) & 17.28 (0.29) &  2 \\
05Jan12 &  932.35 &  17.90 (0.18) & 18.24 (0.07) & 17.83 (0.04) & 17.78 (0.04) & 17.41 (0.04) & 5 \\
10Jan12 &  937.24 &    -- & 18.36 (0.24) & 18.03 (0.24) & 18.14 (0.16) & 17.51 (0.29) &  2 \\
12Jan12 &  939.33 &    -- & 18.49 (0.07) & 18.18 (0.08) & 18.19 (0.07) & 17.54 (0.04) &  3 \\
17Jan12 &  944.23 &  18.65 (0.16) &  -- & 18.47 (0.20) & 18.44 (0.14) & 17.82 (0.16) &  1 \\
17Jan12$^\ast$ &  944.34 &    -- &  -- &  -- & 18.40 (0.28) &  -- & 6 \\ 
18Jan12$^\ast$ &  945.35 &    -- &  -- &  -- & 18.48 (0.18) &  -- & 6 \\
18Jan12 &  945.37 &    -- & 19.00 (0.19) & 18.48 (0.17) &  -- &  -- &  1 \\
21Jan12 &  948.22 &  19.01 (0.41) & 18.97 (0.26) & 18.58 (0.13) & 18.59 (0.16) & 17.95 (0.20) & 1 \\
27Jan12 &  954.24 &    -- & 19.45 (0.15) & 19.08 (0.18) & 18.95 (0.19) & 18.57 (0.21) &  1 \\
30Jan12 &  957.22 &    -- & 19.92 (0.30) & 19.35 (0.24) & 19.34 (0.30) & 18.58 (0.27) &  1 \\
25Jul12 & 1133.67 &   -- & --&  -- & $>$23.16 &  $>$23.08 & 6 \\
02Aug12 & 1141.55 &   -- & -- & $>$20.44  & -- & -- & 1 \\
21Aug12 & 1160.60 & $>$23.29 & $>$23.85 & $>$24.15 & -- & -- & 7 \\ 
\hline
\end{tabular}
\\
1 = 1.82m Copernico Telescope + AFOSC (Mt. Ekar, Asiago, Italy);
2 = 67/92-cm Schmidt Telescope + SCAM (Mt. Ekar, Asiago, Italy);
3 = 2.2-m Calar Alto Telescope + CAFOS (Calar Alto Obs., Almer\`ia, Spain);
4 = 2.0-m Faulkes Telescope North + EM03 (Haleakala, Hawaii Isl., USA);
5 = 2.0-m Liverpool Telescope + RATCam (La Palma, Canary Isl., Spain);
6  = 3.58-m Telescopio Nazionale Galileo + Dolores (La Palma, Canary Isl., Spain);
7 = 4.2-m William Herschel Telescope + ACAM (La Palma, Canary Isl., Spain).
\end{center}
\label{tab_ph_11hw}
\end{table*}%

\begin{table*}
\caption{Table with the Swift/UVOT band photometry of SNe 2010al and 2011hw. Template subtraction was applied to the UVOT images of both SNe.
Original $u$, $b$, $v$ UVOT magnitudes have been converted to those in the Johnson-Cousins photometric system using the magnitudes of the 
stellar sequences reported in Table \ref{tab_seqstars}.
}
\begin{center}
\small
\begin{tabular}{ccccccccc} \hline\hline
\multicolumn{8}{c}{SN 2010al} \\ \hline
Date & $JD+$2455000 & $uvw2$ & $uvm2$ & $uvw1$ & $u$ & $b$  & $v$  \\ \hline
%   & 2455000 & & & & & & & \\ \hline
23Mar10 & 279.27 & 14.64 (0.14)  &  14.60  (0.07) &  14.58 (0.08)  &   --          & --            &     --        \\
28Mar10 & 284.38 & 16.06 (0.18)  &  15.61  (0.10) &  15.33 (0.08)  &  15.06 (0.09) &  15.98 (0.10) &  15.80 (0.08) \\
30Mar10 & 286.05 & 16.32 (0.20)  &  15.97  (0.18) &  15.56 (0.09)  &  15.22 (0.09) &  16.07 (0.09) &  15.86 (0.12) \\
01Apr10 & 288.06 & 16.68 (0.17)  &  16.52  (0.10) &  15.86 (0.10)  &  15.30 (0.10) &  16.12 (0.10) &  15.81 (0.10) \\  
03Apr10 & 289.87 & 17.13 (0.17)  &  16.93  (0.25) &  16.12 (0.12)  &  15.46 (0.13) &  16.26 (0.10) &  15.89 (0.14) \\ 
05Apr10 & 291.74 & 17.61 (0.18)  &  17.55  (0.18) &  16.73 (0.19)  &  15.72 (0.11) &  16.33 (0.10) &  16.08 (0.14) \\
07Apr10 & 293.72 & 18.01 (0.19)  &  17.85  (0.19) &  16.93 (0.15)  &  15.78 (0.11) &  16.51 (0.11) &  16.20 (0.11) \\ 
09Apr10 & 295.96 & 18.35 (0.33)  &  18.36  (0.36) &  17.48 (0.36)  &  16.15 (0.19) &  16.76 (0.19) &  16.28 (0.21) \\  
11Apr10 & 297.81 & 18.87 (0.34)  &  18.57  (0.46) &  17.86 (0.28)  &  16.58 (0.13) &  16.91 (0.18) &  16.54 (0.17) \\  
13Apr10 & 299.73 & 19.22 (0.48)  &  18.88  (0.28) &  18.18 (0.29)  &  16.82 (0.14) &  17.12 (0.12) &  16.72 (0.19) \\  
17Apr10 & 304.26 & $>$19.53      &  19.31  (0.38) &  18.73 (0.31)  &  17.93 (0.29) &  18.04 (0.20) &  17.53 (0.22) \\ 
19Apr10 & 306.06 & $>$19.48      &  $>$19.30      &  18.95 (0.33)  &  18.15 (0.33) &  18.29 (0.28) &  17.88 (0.30) \\   
23Apr10 & 310.18 & $>$19.44      &  $>$19.31      &  19.26 (0.37)  &    $>$18.49   &  18.92 (0.30) &   $>$17.89    \\  
25Apr10 & 311.75 & $>$19.33      &  $>$19.31      &  $>$18.64      &    $>$18.42   &   $>$18.85    &   $>$17.84    \\  
27Apr10 & 313.55 & $>$19.45      &  $>$19.36      &  $>$18.69      &    $>$18.48   &   $>$18.92    &   $>$17.98    \\  
29Apr10 & 316.03 & $>$19.63      &  $>$19.41      &  $>$18.78      &    $>$18.67   &   $>$19.03    &   $>$18.21    \\  
03May10 & 319.98 & $>$19.46      &  $>$19.37      &  $>$18.82      &    $>$18.59   &   $>$18.81    &   $>$18.14    \\  
05May10 & 322.09 & $>$19.32      &  $>$19.22      &  $>$18.79      &    $>$18.30   &   $>$18.58    &   $>$17.93    \\  
27Feb11 & 588.39 & $>$19.46      &  $>$19.32      &  $>$18.62      &    $>$19.57   &   $>$18.79    &   $>$18.01    \\
\hline\hline
\multicolumn{8}{c}{SN 2011hw} \\ \hline
Date & $JD+$2455000 & $uvw2$ & $uvm2$ & $uvw1$ & $u$ & $b$  & $v$  \\ \hline
%   & 2455000 & & & & & & & \\ \hline
22Nov11 & 888.27 & 17.48 (0.06) & 17.34 (0.07)  & 16.89 (0.07) & 16.46 (0.06) & 17.11 (0.07) & 16.93 (0.09)\\
24Nov11 & 890.45 & 17.70 (0.07) & 17.41 (0.07)  & 16.87 (0.07) & 16.59 (0.07) & 17.15 (0.07) & 16.97 (0.09) \\
26Nov11 & 891.87 & 17.70 (0.07) & 17.42 (0.07)  & 16.90 (0.07) & 16.57 (0.07) & 17.15 (0.07) & 16.93 (0.09)\\
28Nov11 & 893.96 & 17.69 (0.07) & 17.38 (0.07)  & 16.90 (0.07) & 16.58 (0.07) & 17.14 (0.08) & 17.03 (0.10)\\
30Nov11 & 896.17 & 17.51 (0.07) & 17.14 (0.07)  & 16.62 (0.06) & 16.48 (0.06) & 17.03 (0.07) & 17.00 (0.09)\\
14Apr13 & 1397.24& $>$19.81   &  $>$19.67  &  $>$19.46 &    $>$19.00 &   $>$19.02 &   $>$18.34 \\
\hline
\end{tabular}
\end{center}
\label{tab_ph_swift}
\end{table*}%

\bsp

\label{lastpage}

\end{document}